\UseRawInputEncoding
\documentclass[aps,prb,floatfix,superscriptaddress,twocolumn,footinbib]{revtex4-1}
\usepackage{amssymb}
\usepackage{amsmath}
\usepackage{amsfonts}
\usepackage{appendix}
\usepackage{bm}
\usepackage{graphicx}
\usepackage{epsfig}
\usepackage{epstopdf}
\usepackage{balance}
\usepackage[dvipsnames]{xcolor}
\usepackage{calc}
\usepackage{natbib}
\usepackage[colorlinks,
linkcolor=blue,
anchorcolor=blue,
citecolor=blue,
urlcolor=blue]{hyperref}
\usepackage{lipsum}
\usepackage{braket}
\usepackage{gensymb}

\begin{document}
	\title{Mirror symmetry decomposition in  double-twisted multilayer graphene systems}
	\author{Shi-Ping Ding}
	\affiliation{School of Physics and Wuhan National High Magnetic Field Center,
		Huazhong University of Science and Technology, Wuhan 430074,  China}
	\author{Miao Liang}
	\affiliation{School of Physics and Wuhan National High Magnetic Field Center,
		Huazhong University of Science and Technology, Wuhan 430074,  China}
	\author{Zhen Ma}
	\email{mazhen@hust.edu.cn}
	\affiliation{School of Physics and Wuhan National High Magnetic Field Center,
		Huazhong University of Science and Technology, Wuhan 430074,  China}
	\author{Jing-Tao L{\"u}}     
	\email{jtlu@hust.edu.cn}
	\affiliation{School of Physics and Wuhan National High Magnetic Field Center,
		Huazhong University of Science and Technology, Wuhan 430074,  China}
	\author{Jin-Hua Gao}
	\email{jinhua@hust.edu.cn}
	\affiliation{School of Physics and Wuhan National High Magnetic Field Center,
		Huazhong University of Science and Technology, Wuhan 430074,  China}
	\begin{abstract}
		Due to the observed superconductivity, the alternating twisted trilayer graphene (ATTLG) has drawn great research interest very recently, in which three monolayer graphene (MLG) are stacked in alternating twist way. If one or several of the MLG in ATTLG are replaced by a multilayer graphene, we get a double twisted multilayer graphene~(DTMLG). In this work, we theoretically illustrate that, if the DTMLG has a mirror symmetry along z direction like the ATTLG,  there exists a mirror symmetry decomposition (MSD), by which the DTMLG can be exactly decoupled into two subsystems with opposite parity. The two subsystems  are either a twisted multilayer graphene (single twist) or a multilayer graphene, depending on the stacking configuration. Such MSD can give a clear interpretation about   all the novel features of the moir\'{e} band structures of DTMLG, e.g.~the fourfold degenerate flat bands and the enlarged magic angle. Meanwhile, in such DTMLG, the parity becomes a new  degree of freedom of the electrons, so that we can define a parity resolved Chern number for the moir\'{e} flat bands.   More importantly, the MSD implies that all the novel correlated phases in the twisted  multilayer graphene should also exist in the corresponding DTMLGs, since they have the exact same Hamiltonian in form. Specifically, according to the MSD,  we predict that the superconductivity should exist in the  (1+3+1)-DTMLG.  
	\end{abstract}
	\maketitle
	\section{Introduction}
	Alternating twisted trilayer graphene (ATTLG) has drawn great research interest very recently\cite{park2021tunable,Haoscience2021,cao2021pauli, nature2022, liuxiaoxue2022natphy, science2022, Khalaf2019prb, carr2020nanolett, lixiao2019arXiv, Leichao2021PRB, yuan2021china, ledwith2021tb,shin2021prb, xiefang2021prb, lianbiao2021prb, lado2021prl, prx2022, nijun2022prb, senthil2021prb, sarma2021prl, qin2021prl, guinea2021prb, vogl2021prb, 2Dmaterial2022, phonon2022prb, mora2022prr, adma2022, npj2022}. It is mainly because that, except the twisted bilayer graphene (TBG)\cite{cao2018correlated,cao2018unconventional,lu2019superconductors,PhysRevLett.124.076801,andrei2020graphene,polshyn2019,lisi2021,choi2019,sharpe2019,kazmierczak2021strain,choi2019,saito2021isospin,wong2020cascade,oh2021evidence,yoo2019atomic,gadelha2021localization,jiang2019charge,xie2019spectroscopic,liu2021tuning},
 the ATTLG is the second definite moir\'{e} system,  in which robust superconductivity is observed\cite{park2021tunable,Haoscience2021,cao2021pauli, nature2022, liuxiaoxue2022natphy}. The ATTLG has a sandwich structure, where three  monolayer graphene (MLG) are stacked in alternating twist way with two twist angles ($\theta_{12}=-\theta_{23}$). Thus, the ATTLG has a mirror symmetry $M_z$ along the z direction. An immediate consequence of the mirror symmetry is that the Hamiltonian of the ATTLG can be exactly mapped into two subsystems\cite{Khalaf2019prb, carr2020nanolett, lixiao2019arXiv}, namely a TBG and a MLG, which have opposite parity~(i.e.~the eigenvalue of the $M_z$). Interestingly, such decomposition shows that the moir\'{e} interlayer tunneling in the TBG subsystem is scaled by a  factor $\sqrt{2}$,  so that the ATTLG has a larger magic angle $\sqrt{2} \times 1.05^\circ \approx 1.54^\circ$  ($1.05^\circ$ is the magic angle of the TBG)\cite{Khalaf2019prb, carr2020nanolett, lixiao2019arXiv}. Meanwhile, the decomposition also clearly interprets the moir\'{e} band structure of the ATTLG near $E_f$, where  one pair of moir\'{e} flat bands results from the TBG subsystem and the two linear bands  are from the MLG part. It is believed that the intriguing superconductivity in ATTLG arise from the TBG subsystem of the Hamiltonion, since it has the same symmetry as that of TBG\cite{park2021tunable,Haoscience2021,cao2021pauli,nature2022, liuxiaoxue2022natphy}.

	Stimulated by the success of the ATTLG, people expect to find more moir\'{e} structures similar as the ATTLG, which can also host the superconducting phase. Very excitingly, several recent experiments report that the superconductivity will be enhanced in alternating twisted four- and five-layer graphene\cite{ParkNatMa2022,fourlayer2022natmaterial}, which are examples of the alternating twisted multilayer graphene (ATMLG) with $n=4,5$ ($n$ is the total number of layers). The ATMLG means each MLG ($n \geq 3$) is twisted in an alternating way, i.e.~the twist angles are $(\theta,-\theta,\theta,\cdots)$\cite{Khalaf2019prb,ledwith2021tb,jj2022prb,LIU202228}.  Theoretically, it is proved that the ATMLG can be decoupled into $\frac{n}{2}$ ($\frac{n-1}{2}$) copies of TBG if $n$ is even (odd), with an additional MLG for  odd $n$\cite{Khalaf2019prb,ledwith2021tb}.   Note that the mirror symmetry $M_z$ also play a key role for the ATMLG with odd n. 
	
	\begin{figure}[h!]
		\flushleft
		\includegraphics[scale=0.31]{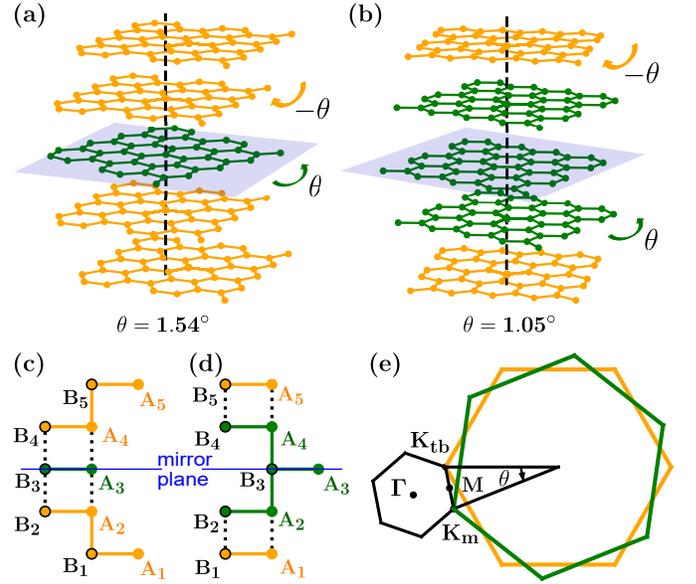}
		\caption{Schematic diagram of the DTMLG with mirror symmetry. (a) is the (2+1+2)-DTMLG and (b) is the (1+3+1)-DTMGL where the twist angles $\theta$ and $-\theta$ have opposite rotation direction. (c) and (d) are the stacking configurations of (a) and (b), respectively. A (B) denotes the sublattice A (B) in each MLG, and the blue lines represent the mirror reflect planes.  (e) is the Brillouin Zone of DTMLG. }
		\label{atomic}
	\end{figure}
	
	The double twisted mutilayer graphene (DTMLG) is another kind of graphene moir\'{e} structures closely related the ATTLG\cite{Liangmiao2022prb,mazhen2023,xie2022alternating,jj2022prb}. The DTMLG is also a sandwich structure like the ATTLG, where three van der Waals (vdW) layers stacked in alternating twist way with two twist angles. But, different from ATTLG,  in DTMLG at least one of the three vdW layers is replaced by a mutilayer graphene\cite{Liangmiao2022prb,mazhen2023}. Note that  we focus on  the alternating twist case in this work, and do not consider the chiral twist situations\cite{mora2019prl,wangke2021prl,zhuziyan2020prl}. 
 Specifically, a general DTMLG  can be denoted as a (X+Y+Z)-DTMLG, where $X$, $Y$ and $Z$ represent the layer number of the bottom, middle and top vdW layers, respectively.
 As shown in Fig.~\ref{atomic}, we plot the structures of two simple DTMLGs as examples, i.e.~(1+3+1)-DTMLG  and (2+1+2)-DTMLG. 
	In (1+3+1)-DTMLG, the middle vdW layer becomes a Bernal-stacked graphene trilayer, while the top and bottom vdW layers are still MLGs.   Meanwhile, the (2+1+2)-DTMLG has bilayer graphene (BLG)  as its top and bottom vdW layers, and the middle vdW layer is MLG. Very interestingly, our previous works have illustrated that the DTMLGs have fascinating moi\'{e} band structures as well\cite{Liangmiao2022prb,mazhen2023}. For example, the (1+3+1)-DTMLG has fourfold degenerate flat bands (for single valley and single spin), twice as much as that in TBG, at the magic angle $\theta \approx 1.05^\circ$\cite{mazhen2023}. And the (2+1+2)-DTMLG has a pair of moir\'{e} flat bands at the magic angle $\theta=1.54^\circ$, coexisting with two parabolic bands near $E_f$\cite{Liangmiao2022prb}. Note that  the DTMLG actually is a large family of moir\'{e} heterostructures, since that  various kinds of multilayer graphene can be chosen  as its vdW layers.

	In this work, we consider a special kind of DTMLG, i.e.~\textit{DTMLGs with mirror symmetry} $M_z$ (see Fig.~\ref{atomic}), where the (1+3+1)-DTMLG and (2+1+2)-DTMLG are the two simplest examples.   We show that, like the ATTLG, there is also a mirror symmetry decomposition (MSD) in such kind of DTMLG, by which the DTMLG can be exactly mapped into 
	two subsystems  with opposite parity. Depending on its own stacking configuration, the two subsystems are either a  twisted multilayer graphene (single twist)  or a multilayer graphene (no twist). 
	
	\begin{figure}[t!]
		\flushleft
		\includegraphics[scale=0.47]{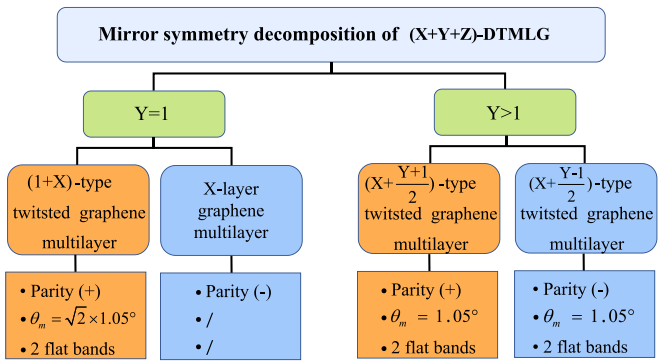}
		\caption{Classification of the mirror symmetry decomposition (MSD) for a general (X+Y+Z)-DTMLG. }\label{fig2}
	\end{figure}
	
  For a general (X+Y+Z)-DTMLG, to satisfy the mirror symmetry $M_z$, there are several apparent requirements: (1) $X=Z$, the thickness of the top and bottom vdW layers must be the same; (2) The total number of graphene layers, i.e.~$X+Y+Z$, should be an odd number, which means that $Y$ has to be odd; (3) The middle vdW layer should intrinsically has mirror symmetry. The (1+3+1)-DTMLG is a typical example with $X=Z=1$ and $Y=3$, where the middle vdW layer, i.e.~a Bernal-stacked (ABA-stacked) graphene trilayer, is mirror symmetric relative to the middle MLG. Actually, the middle MLG of the DTMLG here is always the reflection plane of the $M_z$ symmetry, see Fig.~\ref{atomic}. Our study indicates that, depending on the the middle vdW layer is a MLG (i.e.~$Y=1$) or not,  the MSD can be classified into  two categories: 
	\begin{enumerate}
		\item  When $Y=1$, the DTMLG can be decoupled into a (1+X)-type twisted multilayer graphene with a magic angle $\theta=\sqrt{2} \times 1.05^\circ \approx 1.54^\circ$ (even parity) and a X-layer multilayer graphene (odd parity). The (2+1+2)-DTMLG is an example of this case.
		\item When $Y > 1$, the DTMLG can be decoupled into a ($X+\frac{Y+1}{2}$)-type twisted multilayer graphene (even parity) and a ($X+\frac{Y-1}{2}$)-type twisted multilayer graphene (odd parity). For the two subsystems, the magic angles are both $1.05^\circ$, which means there are fourfold degenerate flat bands at the magic angle.  The (1+3+1)-DTMLG is an example of this situation. 
	\end{enumerate}
	Here, the (1+X)-type twisted multilayer graphene means a MLG and a X-layer multilayer graphene are stacked with a twist angle (single twist). The ($X+\frac{Y \pm 1}{2}$)-type twisted multilayer graphene is  defined in the same way, where $\frac{Y \pm 1}{2}$ is the total  number of the layers of the top vdW layer.  The MSD above is the central results of this work, which is summarized in Fig.~\ref{fig2}.

	Such decomposition will give us  several important insight  into the essential features of the DTMLGs with mirror symmetry: (1) All the intriguing features of the moir\'{e} bands of DTMLG, such as the  different magic angles ($1.54^\circ$ or $1.05^\circ$) and fourfold degenerate flat bands, can be clearly interpreted by this decomposition; (2) The parity becomes a new discrete  degree of freedom of the electrons in DTMLG, and we can define parity resolved 
	Chern numbers for the moir\'{e} flat bands. 
	(3) Such exact mapping strongly implies that the exotic correlation phases observed in twist multilayer graphene (single twist) or multilayer graphene all can exist in the DTMLG systems. 
	(4) Due to the mirror symmetry, the vertical electric field $E_\perp$  acts merely as  hybridization terms between the two subsystems, and is not directly applied on the subsystems themselves. It thus indicates that $E_\perp$ will affect the correlation states of the subsystems in a rather different way. 
	
	The paper is organized as follows. In Sec.~II, we first introduce the general continuum Hamiltonion of the DTMLG. Then, we  use the (2+1+2)-DTMLG and (1+3+1)-DTMLG as examples to illustrate the ideas of the MSD in Sec.~III and IV, respectively. Then, we derive the decomposition formulas for a general DTMLG in Sec.~V. Finally, a short summary is given in Sec.~VI.

	\section{The continuum Hamiltonian of DTMLG}
	Based on the continuum model method\cite{pnas2011,moon2013prb,Koshino_2015,moon2015}, the Hamiltonian of a general (X+Y+Z)-DTMLG for one valley and one spin is
	\begin{equation}\label{totalH}
		\begin{aligned}
			H_{X+Y+Z}=\left(
			\begin{array}{ccc}
				H_X(k_1)&T^\dag_{XY}(r)&0\\
				T_{XY}(r)&H_Y(k_2)&T^\dag_{YZ}(r)\\
				0&T_{YZ}(r)&H_Z(k_3)\\
			\end{array}
			\right)
		\end{aligned}
	\end{equation}
	where the $H_{X,Y,Z}$ are the Hamiltonian of the bottom, middle and top vdW layers, i.e.~the Hamiltonian of multilayer graphene\cite{Liangmiao2022prb,mazhen2023,Ma2020front}.  The Brillouin zone of the DTMLG is given in Fig.~\ref{atomic}, and $\theta_{XY}=-\theta_{YZ}=\theta$ is the twist angle.
	
	$T_{XY}$ ($T_{YZ}$) describes the moir\'{e} hopping between the bottom and middle (middle and top) vdW layers.
	\begin{equation}
		T_{XY}=\sum_{n=0,1,2}T^n_{XY}\cdot{}e^{iq_nr}, 
	\end{equation}
	where $q_{n}=2 k_{D} \sin \left(\frac{\theta}{2}\right) \exp \left(i \frac{2 n \pi}{3}\right)$.  $k_D$ is the magnitude of the BZ corner wave vector of a single vdW layer.

	\begin{equation}
		T_{XY}^{n}=I_{XY} \otimes\left(\begin{array}{cc}
			\omega_{\mathrm{AA}} & \omega_{\mathrm{AB}} e^{i \phi_{n}} \\
			\omega_{\mathrm{AB}} e^{-i \phi_{n}} & \omega_{\mathrm{AA}}
		\end{array}\right)
	\end{equation}
	Here, $I_{XY}$ is a matrix with only one nonzero matrix element. Other parameters are: $\phi_{n}=\textrm{sign}(\theta_{XY}) \frac{2n\pi}{3}$, $\omega_{AA}=0.0797eV$. $T_{YZ}$ is given in a similar way. 
	
	\section{mirror symmetry decomposition of (2+1+2)-DTMLG}    
	Here, we consider the (2+1+2)-DTMLG as the first example\cite{Liangmiao2022prb}, which is shown in Fig.~\ref{atomic} (a) and (c). In order to satisfy the mirror symmetry, the (2+1+2)-DTMLG here refers to a (BA-A-AB)-type structure, see Fig.~\ref{atomic} (c), where the moir\'{e} interfaces are aligned before twisting. We can see that the middle vdW layer is just a MLG, which is also the reflection plane of the mirror symmetry $M_z$. 
	
	According to Eq.\eqref{totalH}, the explicit form of the Hamiltonian of the (2+1+2)-DTMLG is
	\begin{equation}\label{H212}
		H_0=
		\left(
		\begin{array}{ccccc}
			H_1&T^{\dag}_{ 1,2}  &0 &0 &0 \\
			T_{ 1,2}&H_2 &\widetilde{T}^{\dag}_{2,3} &0 &0 \\
			0&  \widetilde{T}_{2,3}&H_3 &\widetilde{T}^{\dag}_{3,4} &0 \\
			0&  0&\widetilde{T}_{3,4}&H_4 &T^{\dag}_{4,5}\\
			0&  0& 0&T_{4,5}&H_5
		\end{array}
		\right).
	\end{equation}
	Here, we use the Bloch waves of each MLG  \{$\ket{A_i}$, $\ket{B_i}$ \} as the basis, where $\ket{A_i}$ is the Bloch wave of the A sublattice of the ith graphene layer.  
	$H_i(\vec{k})=-v_F[R(\theta_i)(\vec{k}-\vec{K}^\xi_i)]\cdot\vec{\sigma}$ is the Hamiltonian of  the \textit{i}th graphene layer, and $R(\theta)$ is the rotation matrix.
	Since that the bottom (top) vdW layer is a BA-type (AB-type) BLG, see Fig.~\ref{atomic} (c), 
	$T_{1,2}$ denotes the interlayer hopping between the 1st and 2nd graphene layers (AB stacked, no twist)
	\begin{equation}
		T_{1,2}= \left(
		\begin{array}{ccc}
			0& t_{\perp}  \\\\
			0& 0 
		\end{array}
		\right).
	\end{equation}
	and $T_{4,5}=T^{\dag}_{1,2}$. $\widetilde{T}_{2,3}$ is the moir\'{e} interlayer hopping,
	\begin{equation}
		\begin{aligned}
			\widetilde{T}_{2,3}&=\sum_{n=0,1,2}\left(
			\begin{array}{cc}
				\omega_{\mathrm{AA}} & \omega_{\mathrm{AB}} e^{i \phi_{n}} \\
				\omega_{\mathrm{AB}} e^{-i \phi_{n}} & \omega_{\mathrm{AA}}
			\end{array}
			\right)e^{i\mathbf{q}_n\cdot\mathbf{r}}
		\end{aligned}
	\end{equation}
	which depends on  the twist angle $\theta$. $\widetilde{T}_{3,4}$ is given in a similar way. 
	
	The idea of MSD in DTMLG is quite like that in the ATTLG \cite{Khalaf2019prb,carr2020nanolett,ledwith2021tb} as well as the ABA-stacked trilayer graphene\cite{koshino2009prb,zhangfan2019pnas}. Namely, we build a new basis which satisfies the mirror symmetry. Based on this new basis, the Hamiltonian \eqref{H212} can be decoupled into two subsystems with opposite parity. Because that the middle graphene layer is the mirror reflection plane, we see that the 1st (2nd) and the 5th (4th) graphene layers are exactly aligned, as shown in Fig.~\ref{atomic} (c). Thus, a natural choice is $\{ \ket{A_{1,5},+}$, $\ket{B_{1,5},+}$, $\ket{A_{2,4},+}$, $\ket{B_{2,4},+}$, $\ket{A_{3}}$, $\ket{B_{3}}$, $\ket{A_{2,4},-}$, $\ket{B_{2,4},-}$, $\ket{A_{1,5},-}$, $\ket{B_{1,5},-} \}$, where 
	\begin{equation}
		\begin{aligned}
			\ket{A_{1,5},+}	&=\frac{1}{\sqrt{2}}(\ket{A_1}+\ket{A_5})\\
			\ket{A_{1,5},-}	&=\frac{1}{\sqrt{2}}(\ket{A_1}-\ket{A_5}).\\
		\end{aligned}
	\end{equation}
	Here, $\ket{A_{1,5},\pm}$ are the two parity resolved basis functions coming  from the A sublattice of the 1st and 5th graphene layers, where  $+$ ($-$) denotes the even (odd) parity. And, $\ket{B_{1,5},\pm}$ are for the B sublattice accordingly. Meanwhile, $\ket{A_{2,4},\pm}$ and $\ket{B_{2,4},\pm}$ correspond to the 2nd and 4th graphene layers, which are defined in the same way. This parity resolved basis gives rise to an unitary transformation,  
	\begin{align}
		U=\frac{1}{\sqrt{2}}
		\left(
		\begin{array}{ccccc}
			I& 0 &0 & 0&I \\
			0&  I& 0& I&0 \\
			0&  0& \sqrt{2}I& 0& 0\\
			0&  I& 0& -I&0 \\
			I& 0& 0& 0&-I
		\end{array}
		\right),
	\end{align}
	where $I$ is a $2 \times 2$ identity matrix. Under this transformation,  $H_{mirr}=U^{-1}H_0U$, where
	\begin{equation}\label{Hmirr}
		\begin{split} 
			H_{mirr}    &= \left( 
			\begin{array}{cccc}
				H_{\textrm{TMBG}} &  0 \\\\
				0&  H_{\textrm{BLG}} 
			\end{array}
			\right )   \\\\\                     
			&=\left(
			\begin{array}{ccccc}
				H_1&T^{\dag}_{ 1,2} &0 &0&0 \\
				T_{ 1,2}&H_2 &\sqrt{2}\widetilde{T}^{\dag}_{2,3} &0&0\\
				0& \sqrt{2}\widetilde{T}_{2,3}&H_3& 0&0 \\
				0& 0&0 &H_2&T_{ 1,2}\\
				0&0&0 &T^{\dag}_{ 1,2}&H_1
			\end{array}
			\right).     
		\end{split}
	\end{equation}
	We see that $H_{mirr}$ is decoupled into two subsystems: $H_{\textrm{TMBG}}$ and $H_{\textrm{BLG}}$.
	Importantly, $H_{\textrm{TMBG}}$ is exactly equivalent to the Hamiltonian of a twisted mono-bilayer graphene (TMBG)\cite{MA202118,jj2020prb,rademaker2020prr,chen2021TMBG,xu2021tmbg,polshyn2020tmbg,he2021tmbg,yin2022tmbg}, except that the moir\'{e} interlayer hopping is multiplied by a factor of $\sqrt{2}$, i.e.~$\sqrt{2}\widetilde{T}_{2,3}$. $H_{\textrm{BLG}}$ is precisely  equal to the Hamiltonian of a BLG.  Meanwhile, according to the parity resolved  basis, $H_{\textrm{TMBG}}$ is of even parity, while the $H_{\textrm{BLG}}$ belongs to odd parity. 
	
	\begin{figure}[t!]
		\flushleft 
		\includegraphics[scale=0.395]{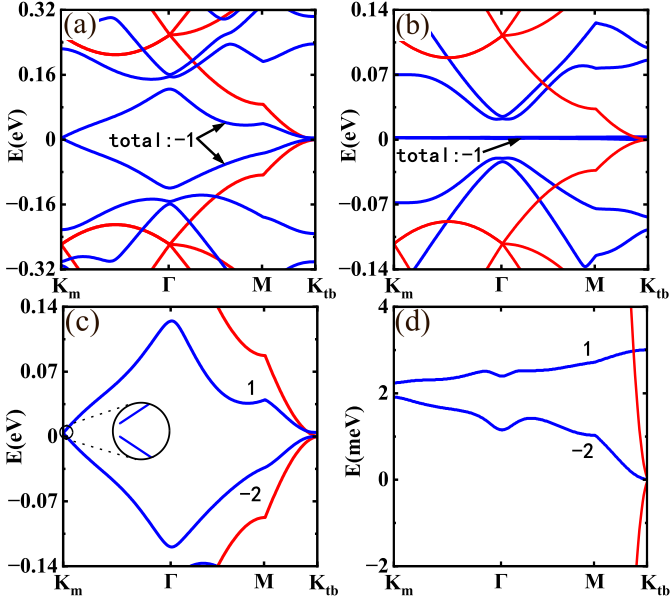}
		\caption{Band structure of the (2+1+2)-DTMLG with mirror symmetry. (a) is for $\theta=2.65^\circ$; (b) is for $\theta=1.54^\circ$, (c) and (d) are enlarged view of the flat bands in (a) and (b), respectively. Blue(red) solid lines represent the band structure of TMBG(BLG), and the black numbers indicate the Chern numbers for the energy bands.}
		\label{band212}
	\end{figure}
	
	Such decomposition clearly interprets the moir\'{e} band structure of the (2+1+2)-DTMLG. It is known that a TMBG has a pair of flat bands at the magic angle $1.05^\circ$, gapped from other high energy bands\cite{MA202118}. However, in $H_{\textrm{TMBG}}$ above, because  the factor $\sqrt{2}$ in moir\'{e} interlayer hopping, it is expected that the magic angle now should be $\sqrt{2} \times 1.05^\circ \approx 1.54^\circ$ like that in ATTLG\cite{Khalaf2019prb,carr2020nanolett}. Meanwhile, due to the $H_{\textrm{BLG}}$, we also expect that there should be a pair of parabolic bands like that in BLG touching  at the $K_{tb}$ point. Then, we plot the energy bands of $H_{mirr}$ in Fig.~\ref{band212} (a) and (b), which is consistent with our former numerical results\cite{Liangmiao2022prb}. The blue (red) solid lines represent the even-parity (odd-parity) bands. In Fig.~\ref{band212} (b), we do find a perfect flat band (even parity) coexisting with a pair of parabolic bands (odd parity) at $\theta=1.54^\circ$, which coincides well with the MSD above. 
	
	The MSD also indicates that the electrons here now have three discrete degree of freedom, i.e.~spin, valley and parity (even or odd).   It thus  becomes possible to define a parity dependent Chern number for the flat bands. 
	As shown in Fig.~\ref{band212} (b),  the two flat bands near $E_f$ come from an equivalent TMBG, i.e.~$H_{\textrm{TMBG}}$, which are of even parity. Meanwhile, the two parabolic bands are of odd parity. 
	So, though the flat bands and the parabolic bands are degenerate at the $K_{tb}$ point, we can study their topological features separately.
	It is known that a TMBG has two topological flat bands, each of which has a non zero valley Chern number\cite{MA202118}.
	And, different from TBG, the two flat bands in TMBG are not degenerate at the Dirac points, due to  the lack of $C_2T$ symmetry. 
	All these of unique features of the flat bands are  reserved in the (2+1+2)-DTMLG, because of the exactly mapping between $H_{\textrm{TMBG}}$ and  a real TMBG. Fig.~\ref{band212} (c) and (d) are the enlarged view of the flat bands in Fig.~\ref{band212} (a) and (b), respectively. 
	Though the magic angle now is larger, the two flat bands have nonzero valley Chern numbers, namely, that of the upper (lower) flat band is 1 ($-2$), which are the same as that in a real TMBG at the magic angle $1.05^\circ$\cite{MA202118}.   
	As for the $H_{\textrm{BLG}}$, we can exclusively calculate the Berry curvature of the parabolic bands, which is same as that of the BLG. 
	
	\begin{figure}[t!]
		\flushleft 
		\includegraphics[scale=0.395]{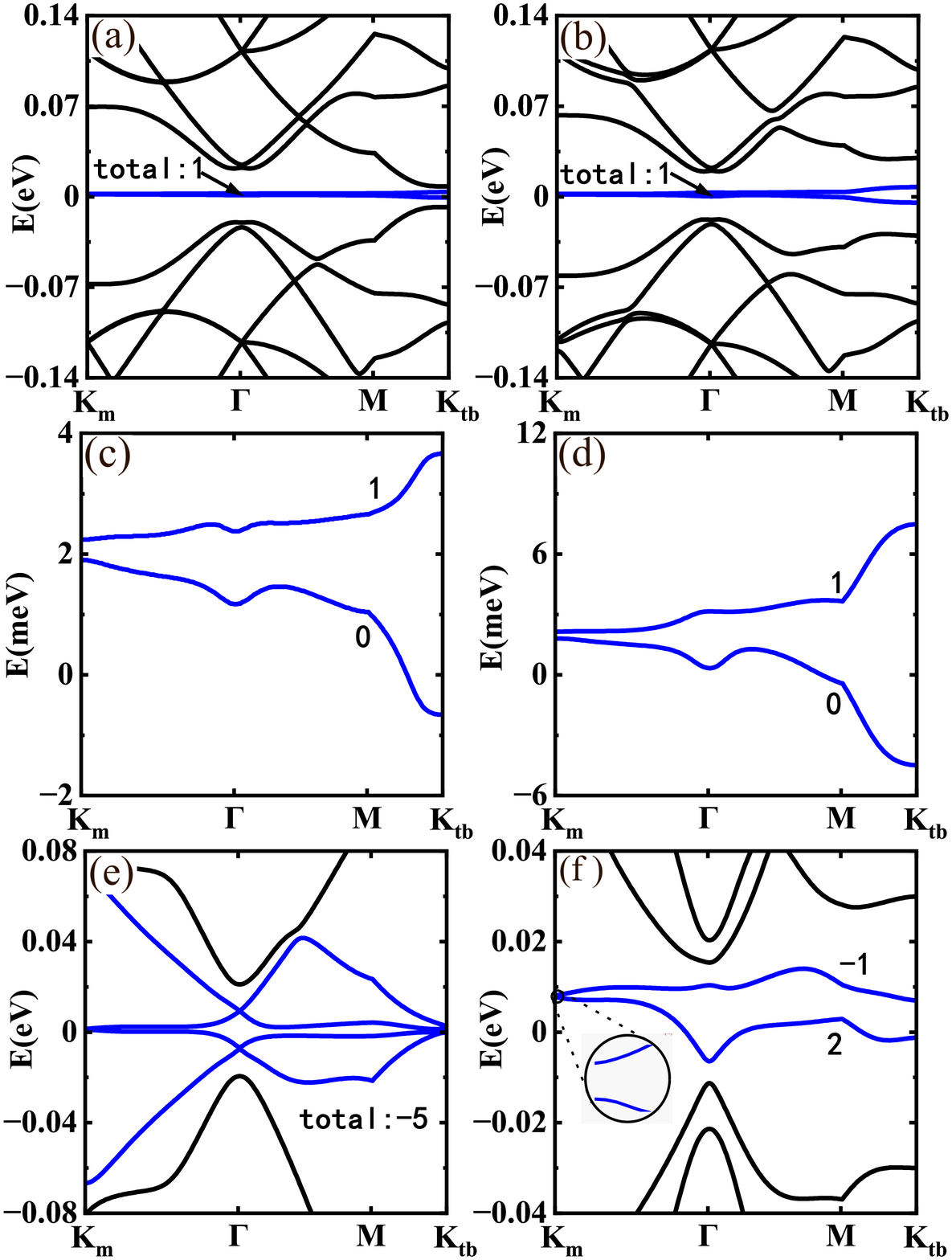}
		\caption{Band structures of the (2+1+2)-DTMLG at $\theta=1.54^\circ$ in the presence of $E_\perp$ or lateral shift. (a) $V=4$ meV. (b) $V=15$ meV.  (c) and (d) are the enlarged view of the flat bands in (a) and (b), respectively. 
  (e) and (f) show the energy bands of the $H_{BA+A+CA}$ after sliding, where (e) $V=0$ meV and (f) $V=15$ meV. }
		\label{bandE212}
	\end{figure}
	
	Now, we discuss the influence of the perpendicular electric field $E_\perp$. Since that $E_\perp$ does not commute with $M_z$, it actually does not directly act  on $H_{\textrm{TMBG}}$ and $H_{\textrm{BLG}}$, but  works as the hybridization terms between the two subsystems\cite{ledwith2021tb}. In the parity resolved basis, the Hamiltonian of $E_{\perp}$ is 
	\begin{equation}\label{Efield1}
		H_V=\left(
		\begin{array}{cc}
			0&D\\
			D^\dag&0
		\end{array}
		\right)
	\end{equation}
	\begin{equation}\label{Efield2}
		D=\left(
		\begin{array}{cc}
			0&2VI\\
			VI&0\\
			0&0
		\end{array}
		\right),
	\end{equation}
	where $D$ is the hybridization matrix, $V$ is the potential difference between adjacent layers and $I$ is a $2 \times 2$ identity matrix. Here, we assume that the $E_\perp$ induced potential will uniformly distributed among the graphene layers. Clearly, $H_V$  indicates that the $E_\perp$ has different ways to couple with a real TMBG and with the subsystem $H_{\textrm{TMBG}}$, though they have the same Hamiltonian in form. In a real TMBG, $E_\perp$ will shift the energy of the states at $K_m$ and $K_{tb}$ towards opposite directions, separate the two flat bands and change their valley Chern number\cite{MA202118}. 
	However,  in the (2+1+2)-DTMLG here, $E_\perp$ merely works as the hybridization between the  two subsystems and thus has a completely different influences.
	First of all,  due to $E_\perp$, the parity is no longer a good quantum number now. 
	In Fig.~\ref{bandE212} (a) and (b), we plot the moir\'{e} bands of the (2+1+2)-DTMLG in the presence of $E_\perp$.
	We see that, once an $E_\perp$ is applied, the degeneracy of the parabolic bands at $K_{tb}$ point is broken immediately, and the central two flat bands are isolated from other bands by an obvious gap. Meanwhile, $E_\perp$ will  induce a hybridization between the flat bands and other odd parity bands, which changes their Chern numbers at once. We zoom in the flat bands in Fig.~\ref{bandE212} (c) and (d), which shows that the Chern number of lower flat band is changed from  $-2$ to $0$. It should be noted that $E_\perp$ has little effects on the states near the $K_m$ point. The gap between the two flat bands at $K_m$ is still very tiny, even a large $E_\perp$ is applied. 
	Another distinct feature is that the effects of $E_\perp$ is irrelative to the direction of $E_\perp$ because the mirror symmetry of (2+1+2)-DTMLG. It  is different from the case of TMLG, where the top MLG and bottom BLG are asymmetric\cite{MA202118}.

	The lateral shift of the moir\'{e} interface plays an importnat role in the ATTLG and DTMLG systems, which not only breaks the mirror symmetry, but also significantly modify the band structures\cite{Leichao2021PRB,shin2021prb}. So far, accurately control of the lateral shift  in moir\'{e} heterostructure is still impossible. Here, the MSD above also can give some insight about the lateral shift. As shown in Fig.~\ref{atomic}, we always assume that the graphene layers of a moir\'{e} interface are aligned when $\theta=0$.  For example, the (2+1+2)-DTMLG above has a (BA+A+AB)-type configuration. If the top vdW layer is shifted, we then get a (BA+A+CA)-DTMLG. To get the Hamiltonian of (BA+A+CA)-DTMLG, we only need to replace the $\widetilde{T}_{34}$ in Eq.~\eqref{H212} by $\widetilde{T}'_{34}$ accordingly,
	\begin{equation}
		\widetilde{T}'^\dag_{34}=\sum_{n=0,1,2}\left(
		\begin{array}{cc}
			\omega_{\mathrm{AA}} & \omega_{\mathrm{AB}} e^{i \phi_{n}} \\
			\omega_{\mathrm{AB}} e^{-i \phi_{n}} & \omega_{\mathrm{AA}}
		\end{array}
		\right)e^{i(\mathbf{q}_n\cdot\mathbf{r}+\phi_n)}
	\end{equation}
	With the same unitary transformation, we get $H_{\textrm{BA+A+CA}} = H_{mirr}+H_{shift}$, where all the effects of lateral shift are attributed to $H_{shift}$,
	\begin{equation}
		H_{shift}=\left(
		\begin{array}{ccccc}
			0&0&0&0&0  \\
			0&0&-\widetilde{T}'^{\dag}&0&0  \\
			0&-\widetilde{T}'&0&\widetilde{T}'&0  \\
			0&0&\widetilde{T}'^{\dag}&0&0  \\
			0&0&0&0&0  
		\end{array}
		\right).
	\end{equation}
	Here, $\widetilde{T}'=\frac{\sqrt{2}}{2}(\widetilde{T}_{2,3}-\widetilde{T'}^{\dag}_{3,4})$. In the mirror symmetric (BA+A+AB)-configuration,  $\widetilde{T}_{2,3}=\widetilde{T}^\dag_{3,4}$ and thus $\widetilde{T}'=0$. But for the (BA+A+CA)-configuration after sliding, $\widetilde{T}_{2,3} \neq \widetilde{T'}^\dag_{3,4}$, so that $\widetilde{T}'$ becomes nonzero. In fact, the $\widetilde{T}'$ represents the effects of lateral shift. It not only appears in the hybridization terms to break the mirror symmetry, but also occurs in the even-parity block. Therefore, the influence of lateral shift is more complicated than that of the $E_\perp$.  In Fig.~\ref{bandE212} (e), we plot the band structure of the (BA+A+CA)-configuration after sliding. We see that the two flat bands and two parabolic bands are heavily hybridized, but we still can get a large DOS at $E_f$. Meanwhile, 
	the central four bands near $E_f$ are isolated from other high energy bands, and their total valley Chern number  is $-5$. Very interestingly, when we apply an electric field $E_\perp$ on the (BA+A+CA)-configuration,  we eventually get a pair of flat bands isolated from other bands with nonzero valley Chern number, quite like that in the (BA+A+AB)-configuration, see Fig.~\ref{bandE212} (f).  
	
	\section{mirror symmetry decomposition of (1+3+1)-DTMLG}
	Then, we discuss the case of (1+3+1)-DTMLG\cite{mazhen2023}. Compared with the (2+1+2)-DTMLG above, the middle vdW layer now becomes a Bernal stacked trilayer graphene, but the middle MLG is still mirror reflection plane (see Fig.~\ref{atomic}).  
	
	Based on Eq.~\eqref{totalH}, The Hamiltonian of (1+3+1)-DTMLG is 
	\begin{equation}\label{H131}
		H_0=
		\left(
		\begin{array}{ccccc}
			H_1&\widetilde{T}^{\dag}_{12}  &0 &0 &0 \\
			\widetilde{T}_{12}&H_2  & {T}^{\dag}_{23}&0 &0 \\
			0&  {T}_{23}&H_3 & {T}^{\dag}_{34} &0 \\
			0&  0& {T}_{34} &H_4 & \widetilde{T}^{\dag}_{45}\\
			0&  0& 0& \widetilde{T}_{45}&H_5
		\end{array}
		\right).
	\end{equation}
	Note that the moir\'{e} interlayer hopping now is $\widetilde{T}_{1,2}$ ( $\widetilde{T}_{4,5}$), which exists between the first (4th) and second (5th) graphene layers.
	
	Since the middle graphene layer is still the mirror reflection plane,  we can still use the parity resolved basis in last section to do the MSD. After the unitary transformation, we get
	
	\begin{equation}\label{Hmirr-131}
		\begin{split} 
			H_{mirr}    &= \left( 
			\begin{array}{cccc}
				H'_{\textrm{TMBG}} &  0 \\\\
				0&  H_{\textrm{TBG}} 
			\end{array}
			\right )   \\\\\    
			&=\left(
			\begin{array}{ccccc}
				H_1&\widetilde{T}^{\dag}_{ 1,2} &0 &0&0 \\
				\widetilde{T}_{ 1,2}&H_2 &\sqrt{2} {T}^{\dag}_{2,3} &0&0\\
				0& \sqrt{2} {T}_{2,3}&H_3& 0&0 \\
				0& 0&0 &H_2&\widetilde{T}_{ 1,2}\\
				0&0&0 &\widetilde{T}^{\dag}_{ 1,2}&H_1
			\end{array}
			\right).     
		\end{split}
	\end{equation}
	Here, we see that the (1+3+1)-DTMLG is mapped into two subsystems: an equivalent TMBG $H'_{\textrm{TMBG}}$ (even parity) and an equivalent TBG $H_{\textrm{TBG}}$ (odd parity). 
	The  two subsystems here are different from that of the  (2+1+2)-DTMLG. First,  the even-parity part $H'_{\textrm{TMBG}}$ in Eq.~\eqref{Hmirr-131} here is distinct from the $H_{\textrm{TMBG}}$ in Eq.~\eqref{H212}.
	It is because  that the $\sqrt{2}$ factor is now at the interlayer hopping without twist, instead of the moir\'{e} interlayer hopping like in Eq.~\eqref{H212}.
	Therefore, in (2+1+2)-DTMLG, the magic angle of  the subsystem $H_{\textrm{TMBG}}$ is $\sqrt{2}\times 1.05^\circ$, but the magic angle of the $H'_{\textrm{TMBG}}$  in the (1+3+1)-DTMLG  is still $1.05^\circ$. Further numerical calculations show that the enlarged term $\sqrt{2} T_{2,3}$ has little influence on the bands near $E_f$.
	Second, the odd-parity block $H_{\textrm{TBG}}$ now has a moir\'{e} interface and becomes an equivalent TBG, which thus will also give rise to a pair of flat bands at the magic angle $1.05^\circ$. 
	So, in (1+3+1)-DTMLG, we get two pairs of flat bands at the magic angle $1.05^\circ$, where one pair is of even parity and the other is of odd parity. The MSD here  give a clear interpretation about the fourfold degenerate flat bands first reported in our previous work\cite{ma2021doubled}. Interestingly, the alternating twisted multilayer graphene also can be decoupled into two copies of TBGs, but their magic angles are different\cite{Khalaf2019prb,ledwith2021tb}. In that case, we can not get two pairs of flat bands at the same time\cite{ParkNatMa2022,fourlayer2022natmaterial}.   
	
	\begin{figure}[t!]
		\flushleft 
		\includegraphics[scale=0.39]{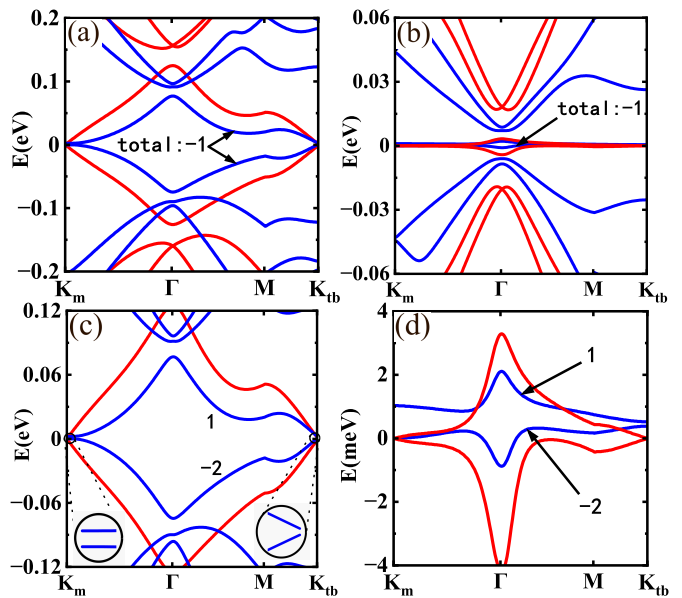}
		\caption{Band structures of the (1+3+1)-DTMLG with mirror symmetry. (a) $\theta=2.0^\circ$, (b) $\theta=1.05^\circ$. (c) and (d) are the enlarged view of the flat bands in (a) and (b), respectively. Blue (red) solid lines represent the band structure of TMBG (BLG), and  the black numbers indicate the Chern numbers for the energy bands.}
		\label{band131}
	\end{figure}
	
	In Fig.~\ref{band131} (a) and (b), we plot the energy bands of the (1+3+1)-DTMLG at two different twisted angles, where blue (red) lines represent the even-parity (odd-parity) bands from $H'_{\textrm{TMBG}}$ ($H_{\textrm{TBG}}$). At $1.05^\circ$, we do observe the fourfold degenerate flat bands, see Fig.~\ref{band131} (b).  Similar as the (2+1+2)-DTMLG, we can also define a parity dependent valley Chern number for the fourfold flat bands.  In Fig.~\ref{band131} (c) and (d), we zoom in the flat bands. As we expected, the two flat bands from $H'_{\textrm{TMBG}}$ (blue lines) have tiny gaps at the $K_m$ and $K_{tb}$ points, which is the typical feature of the flat bands in TMBG\cite{MA202118}. Meanwhile, such two flat bands have nonzero valley Chern number as well, which is also the same as that in TMBG. As for the two flat bands from $H_{\textrm{TBG}}$ (red lines), they are topological trivial and   degenerate at the $K_m$ and $K_{tp}$ points, which is the same as that in TBG. 
	
	\begin{figure}[t!]
		\flushleft 
		\includegraphics[scale=0.39]{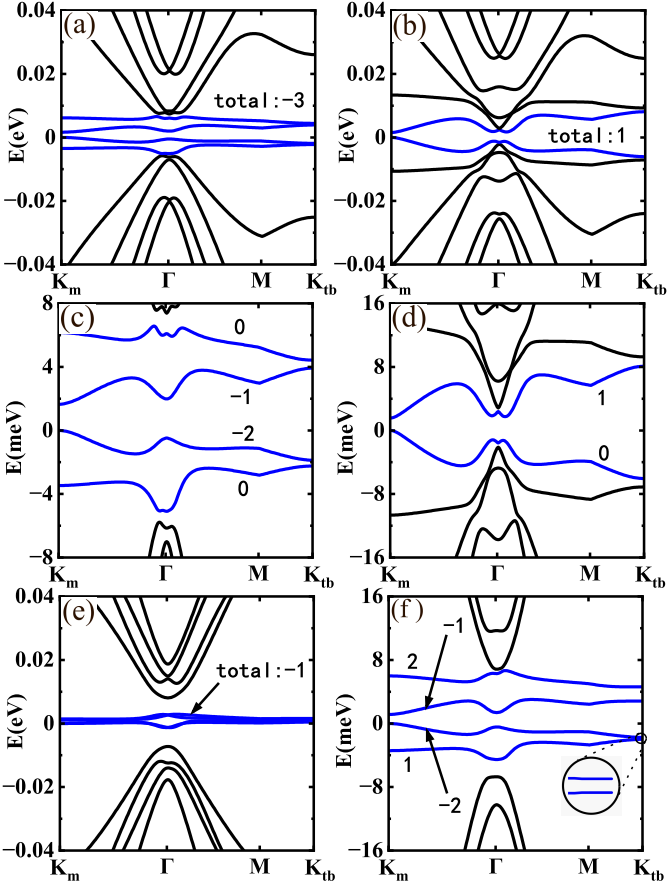}
		\caption{Band structures of the (1+3+1)-DTMLG at $\theta=1.05^\circ$ in the presence of $E_\perp$ or lateral shift. (a) $V=4$ meV, (b) $V=10$ meV.  (c) and (d) are the enlarged view of the flat bands in (a) and (b), respectively. 
   (e) and (f) show the energy bands of the $H_{A+ABA+C}$ after sliding, where (e) $V=0$ meV and (f) $V=4$ meV.}
		\label{bandV}
	\end{figure}
	
	In the parity resolved basis, the Hamiltonian of $E_\perp$, i.e.~$H_V$, has the same form as that of the (2+1+2)-DTMLG, see Eq.~\eqref{Efield1}. 
	$E_\perp$ breaks the mirror symmetry and acts as hybridization terms between the even- and odd-parity blocks, which is similar as that in (2+1+2)-DTMLG. 
	In Fig.~\ref{bandV} (a) and (b), we plot the moir\'{e} bands of the (1+3+1)-DTMLG at $1.05^\circ$ with $V=4$ meV and $V=10$ meV, respectively. And we zoom in flat bands in Fig.~\ref{bandV} (c) and (d).
	Though the (1+3+1)-DTMLG is decoupled into an equivalent TMBG  and an equivalent TBG, the influence of the $E_\perp$ is rather different from that in  real TMBG and TBG.  First, as mentioned above, parity is no longer a good quantum number once $E_{\perp}$ is applied. Meanwhile, the hybridization breaks the flat band degeneracy of the $H_{\textrm{TBG}}$ at the $K_{tb}$ point. In contrast, in a real TBG, we know that  electric field can not lift the degeneracy of flat bands at the Dirac points. 
	Due to the mirror symmetry,  the moir\'{e} bands in the (1+3+1)-DTMLG are unrelated to the direction of $E_\perp$ as well, i.e., $\pm V$ give rise to the same moir\'{e} band structures. Furthermore,  $E_\perp$ also can change the Chern number of the flat bands, as shown in Fig.~\ref{bandV} (c) and (d).  Fig.~\ref{bandV} (b) and (d) further indicate that, when V is large enough, the outermost two flat bands will overlap with the high energy bands in energy. 
	
	We then discuss the effects of the lateral shift. 
	Here, with a lateral shift, we actually get a (A+ABA+C)-configuration, in which  the top vdW layer is shifted. In the parity resolved basis, the effects of the lateral shift can be attributed to a shift term $H_{shift}$
	\begin{equation}\label{Hshift131}
		H_{shift}=\frac{1}{\sqrt{2}}\times
		\left(
		\begin{array}{ccccc}
			0&-\widetilde{T}'^\dag&0&\widetilde{T}'^\dag&0  \\
			-\widetilde{T}'&0&0&0&\widetilde{T}'\\
			0&0&0&0&0  \\
			\widetilde{T}'&0&0&0&-\widetilde{T}'  \\
			0&\widetilde{T}'^\dag&0&-\widetilde{T}'^\dag&0  \\
		\end{array}
		\right),
	\end{equation}
	where $H_{A+ABA+C}=H_{mirr}+H_{shift}$ and  $H_{mirr}$ is the Hamiltonian of the  (A+ABA+A)-configuration. 
	From $H_{shift}$, we see that the lateral shift will not only give rise to a hybridization between the even- and odd-parity blocks, but also have direct effects  on the two subsystems.
	In Fig.~\ref{bandV} (e) and (f), we plot the moir\'{e} bands of the (A+ABA+C)-DTMLG at the magic angle, with $V=0$ and $V=4$ meV respectively. 
	Though the high energy bands are changed, we still get four nearly degenerate flat bands and  $E_\perp$ also can lift the degenerate of the four flat bands. 
	
	\section{mirror symmetry decomposition of a general DTMLG}
	Similar as the two examples above, we actually can give a simple decomposition scheme for a general DTMLG with mirror symmetry. As mentioned above, we consider a general (X+Y+X)-DTMLG with mirror symmetry. For convenience,  we first rewrite the Hamiltonian of Eq.~\eqref{totalH}, 
	\begin{equation}\label{Hgeneral}
		H_0=\left(
		\begin{array}{ccc}
			H_{down}&T^\dag_{dm}&0\\
			T_{dm}&H_{N}&T^\dag_{um}\\
			0&T_{um}&H_{up}
		\end{array}
		\right).
	\end{equation}
	Here,  we define $N=X+\frac{Y+1}{2}$, so that the middle MLG, i.e.~the mirror reflection plane, is just the Nth graphene layer from bottom to up.  $H_{N}$ here is the Hamiltonian of the middle MLG. That we treat the the middle MLG individually here is to facilitate constructing a proper basis.
	Then, all the graphene layers above (below) the middle graphene layer is described by $H_{up}$ ($H_{down}$), which is a $(2X+Y-1) \times (2X+Y-1)$ matrix.  $T_{um}$ ($T_{dm}$) denote the hybridization between $H_{up}$ ($H_{down}$) and $H_N$. 
	
	Considering the position of the moir\'{e} interlayer hopping, there are two different cases depending on whether $Y=1$ or not.   $Y=1$ means that the middle vdW layer is just a MLG, where the (2+1+2)-DTMLG is the simplest example. 
	In this situation, the moir\'{e} interlayer hopping is in $T_{um}$ and $T_{dm}$, where 
	\begin{equation}\label{tdm1}
		T_{dm}=\left(
		\begin{array}{cccc}
			0&\cdots&0&\widetilde{T}_{N,N-1}
		\end{array}
		\right)
	\end{equation}
	\begin{equation}\label{tum1}
		T^{\dag}_{um}=\left(
		\begin{array}{cccc}
			\widetilde{T}^{\dag}_{N,N+1}&0&\cdots&0
		\end{array}
		\right)
	\end{equation}\par  
	and  
	\begin{equation}\label{Hdown1}
		H_{down}=\left(
		\begin{array}{cccc}
			H_1&T^\dag_{1,2}&\cdots&0\\
			T_{1,2}&H_2&\cdots&\vdots\\
			\vdots&\ddots&\ddots&\vdots\\
			0&\cdots&T_{N-2,N-1}&H_{N-1}
		\end{array}
		\right)
	\end{equation}
	\begin{equation}\label{Hupm1}
		H_{up}=\left(
		\begin{array}{cccc}
			H_{N+1}&T^\dag_{N+1,N+2}&\cdots&0\\
			\vdots&\ddots&\ddots&\vdots\\
			\vdots&\cdots&H_{L-1}&T^\dag_{L-1,L}\\
			0&\cdots&T_{L-1,L}&H_{L}
		\end{array}
		\right).
	\end{equation}
	For convenience, we define $L= 2X + Y$  as the total layer number of the DTMLG.  
	When $Y > 1$, e.g. (1+3+1)-DTMLG, the moir\'{e} interlayer hopping is in $H_{up}$ and $H_{down}$. So, we have
	\begin{equation}\label{tdm2}
		T_{dm}=\left(
		\begin{array}{cccc}
			0&\cdots&0&T_{N,N-1}
		\end{array}
		\right)
	\end{equation}
	\begin{equation}\label{tum2}
		T^{\dag}_{um}=\left(
		\begin{array}{cccc}
			T^{\dag}_{N,N+1}&0&\cdots&0
		\end{array}
		\right)
	\end{equation}\par  
	and now  
	
	\begin{equation}\label{Hdown2}
		\begin{aligned}
			&\quad H_{down}=\\
			&\left(
			\begin{array}{ccccc}
				H_1&T^\dag_{1,2}&\cdots&\cdots&0\\
				\vdots&\ddots&\ddots&\ddots&\vdots\\
				\vdots&T_{X-1,X}&H_{X}&\widetilde{T}^\dag_{X,X+1}&\vdots\\
				\vdots&\ddots&\ddots&\ddots&\vdots\\
				0&\cdots&\cdots&T_{N-2,N-1}&H_{N-1}
			\end{array}
			\right)
		\end{aligned}
	\end{equation}
	
	\begin{equation}\label{Hup2}
		\begin{aligned}
			&\quad H_{up}=\\
			&\left(
			\begin{array}{ccccc}
				H_{N+1}&T^\dag_{N+1,N+2}&\cdots&\cdots&0\\
				\vdots&\ddots&\ddots&\ddots&\vdots\\
				\vdots&T_{X+Y-1,X+Y}&H_{X+Y}&\widetilde{T}^\dag_{X+Y,X+Y+1}&\vdots\\
				\vdots&\ddots&\ddots&\ddots&\vdots\\
				0&\cdots&\cdots&T_{L-1,L}&H_{L}
			\end{array}
			\right).
		\end{aligned}
	\end{equation}
	
	The moir\'{e} interlayer hopping now are  the terms $\widetilde{T}_{X,X+1}$ and  $\widetilde{T}_{X+Y, X+Y+1}$. The two cases above will give  different results after the MSD. 
	
	Then, we can construct a set of parity resolved basis,   
	\begin{equation}\label{basis-mirror-general}
		\begin{aligned}
			\ket{A_{i,L+1-i},+}&=\frac{1}{\sqrt{2}}(\ket{A_i}+\ket{A_{L+1-i}})\\
			\ket{A_{i,L+1-i},-}&=\frac{1}{\sqrt{2}}(\ket{A_i}-\ket{A_{L+1-i}})\\
			\ket{A_{N},+}&=\ket{A_{N}} \\
			\ket{B_{N},+}&=\ket{B_{N}}. 
		\end{aligned}
	\end{equation}
	where $i \in \{1,2,\ldots,N-1\}$ and $\pm$ denote the parity.  
	For example, when $i=1$, we get $\ket{A_{1,L},+} =\frac{1}{\sqrt{2}}(\ket{A_1}+\ket{A_{L}})$, which is obviously of even parity considering the mirror symmetry. Similarly, the wave function  $\ket{A_{1,L},-}$ is of odd parity.  $\ket{A_{N},+}$ and  $\ket{B_{N},+}$ are the wave functions of the middle graphene layer, which should be of even parity. 
	With the definitions above, we get a set of parity resolved basis, i.e.~$\{\ket{+},\ket{-}\}$, where the even parity basis is $\ket{+}$ $=$ $\{\ket{A_{1,L},+}$, $\ket{B_{1,L},+}$, $\ket{A_{2,L-1},+}$, $\ket{B_{2,L-1},+}$, $\cdots$, $\ket{A_{N-1,N+1},+}$, $\ket{B_{N-1,N+1},+}$, $\ket{A_{N},+}$, $\ket{B_{N},+}\}$ and the odd parity basis is $\ket{-}$ $=$ $\{\ket{A_{N-1,N+1},-}$, $\ket{B_{N-1,N+1},-}$, $\cdots$, $\ket{A_{2,L-1},-}$, $\ket{B_{2,L-1},-}$, $\ket{A_{1,L},-}$, $\ket{B_{1,L},-}\}$.

	In such basis, the $H_0$ in Eq.~\eqref{Hgeneral} can be decoupled by an  unitary transformation, 
	\begin{equation}\label{Hmirr-general}
		H_{mirr}=U^{-1}H_0 U = \left(
		\begin{array}{ccc}
			H_{+}&0 \\
			0 & H_{-}
		\end{array}
		\right).
	\end{equation}
	The $H_{+}$ ($H_{-}$) is the even (odd) parity block
	\begin{equation}\label{H+}
		\begin{aligned}
			H_{+}&= \left(
			\begin{array}{ccc}
				H_{down}&\sqrt{2}T^{\dag}_{dm}\\
				\sqrt{2}T_{dm}&H_{N}
			\end{array}
			\right)\\
			H_{-}&=H_{up},
		\end{aligned}
	\end{equation}
	The detailed derivation is given in the Appendix.
	Note that, though the $H_{down}$ and $H_{up}$ in $H_{mirr}$ above have the same form as that in Eq.~\eqref{Hdown1}, \eqref{Hupm1},  \eqref{Hdown2} and \eqref{Hup2},  they actually correspond to the parity resolved basis, instead of the original Bloch wave basis. 
	Now, we see that the DTMLG with mirror symmetry can always be decoupled into two subsystems with opposite parity, i.e.~$H_{+}$ and $H_{-}$. As mentioned above, depending on whether the middle vdW layer is a MLG or not, there are two distinct situations in which the $H_{+}$ and  $H_{-}$ correspond to different concrete systems.  The reason is where the moir\'{e} interlayer hopping is. 
	
	When $Y=1$, the moir\'{e} interlayer hopping is in $T_{dm}$ and $T_{um}$, as shown in Eq.~\eqref{tdm1} and \eqref{tum1}. So, $H_{-}=H_{up}$ does not have moir\'{e} interlayer hopping, see Eq.~\eqref{Hupm1}. Namely, $H_{-}$ is just an equivalent multilayer graphene, which has the same stacking order as the top  vdW layer. For example, in (2+1+2)-DTMLG, $H_{-}$ is just a bilayer graphene~(AB-stacking). Meanwhile, from Eq.~\eqref{H+}, \eqref{tdm1} and \eqref{tum1}, we can see that $H_{+}$ now describes a (1+X)-type twisted multilayer graphene (single twist) \cite{Ma2020front,liujianpengprx2019,zhang2020chiral,liuchengcheng2021prb} with a $\sqrt{2}$ scaled moir\'{e} interlayer hopping. It implies that $H_{+}$ should have a pair of flat bands at the magic angle $\sqrt{2} \times 1.05^\circ$. This is also in accordance with the (2+1+2)-DTMLG, where $H_{+}$ is equivalent to a  TMBG with a  $\sqrt{2} \times 1.05^\circ$ magic angle.      
	
	When $Y > 1$, the moir\'{e} interlayer hopping is in $H_{down}$ and $H_{up}$, as shown in Eq.~\eqref{Hdown2} and \eqref{Hup2}. Thus, $H_{-}=H_{up}$ now correspond to a $(X+\frac{Y-1}{2})$-type twisted multilayer graphene, see Eq.~\eqref{Hup2} and definition of $H_{up}$. An example is the (1+3+1)-DTMLG with $Y=3$ and $X=1$. In this example, $H_{-}$ is equal to a  TBG, i.e.~(1+1)-type, which coincides with the general decomposition formula above. Then, we discuss $H_{+}$. From Eq.~\eqref{H+}, \eqref{tdm2} and \eqref{tum2}, we see that the hopping between $H_{N}$ and $H_{down}$ is a normal interlayer hopping between graphene layers, instead of the moir\'{e} one. It implies that 
	$H_+$ now describes a $(X+\frac{Y+1}{2})$-type twisted multilayer graphene with a single twist angle.  In the example of (1+3+1)-DTMLG,  $H_{+}$ thus becomes a TMBG, i.e.~(1+2)-type. Here, since the $\sqrt{2}$ factor is not at the moir\'{e} interlayer hopping, both $H_-$ and $H_+$ should have a pair of flat bands at the magic angle $1.05^\circ$.  Because that the moir\'{e} band structures of twisted multilayer graphene is known\cite{Ma2020front,liujianpengprx2019,zhang2020chiral,liuchengcheng2021prb}, the band structure of such DTMLG with mirror symmetry can be well understood from the MSD, namely, from the behaviors of $H_+$ and $H_-$. 
	
	With parity resolved basis, the perpendicular electric field can be described as hybridization terms between $H_+$ and $H_-$, since it breaks the mirror symmetry. 
	The Hamiltonian of $E_\perp$ is 
	\begin{equation}\label{H-general-V}
		H_V=\left(
		\begin{array}{cc}
			0&D\\
			D^\dag&0
		\end{array}
		\right)
	\end{equation}
	where $H=H_{mirr}+H_V$. $D$ is now a $(2X+Y+1) \times  (2X+Y-1)$ matrix 
	\begin{equation}\label{V-general}
		D=\left(
		\begin{array}{c}
			\Delta \\
			0
		\end{array}
		\right)
	\end{equation}
	and $\Delta$ is a $(X+\frac{Y-1}{2})\times(X+\frac{Y-1}{2})$ block matrix, where each block $\Delta_{i,j}=V(N-i)I\delta_{i,N-j}$ is $2 \times 2$ matrix. As defined before,  $V$ is the potential difference between adjacent graphene layers.
	Clearly, the $H_V$ indicates that $E_\perp$ will not directly act on the $H_+$ and $H_-$, but works as the hybridization between $H_+$ and $H_-$.  
	
	In short, according to the discussions above, the MSD can be summarized in a schematic in Fig.~\ref{fig2}.

	\section{Summary}
	We theoretically reveal that a general (X+Y+Z)-DTMLG with mirror symmetry can be exactly decoupled into two subsystems, i.e., $H_+$ and $H_-$, with opposite parity, by constructing a parity resolved basis. Such MSD can be classified into to two categories, depending on whether the middle vdW layer is a MLG ($Y=1$) or not. 
	
	When $Y=1$, the DTMLG is mapped into a (1+X)-type twisted multilayer graphene (single twist, even parity) and a X-layer graphene multialyer (no twist, odd parity). Since the moir\'{e} interlayer hopping is scaled by a factor of $\sqrt{2}$, the (1+X)-twisted multilayer graphene has a enlarged magic angle $\sqrt{2} \times 1.05^\circ$. The (2+1+2)-DTMLG is the simplest example for this case, which can be decoupled into an equivalent TMBG  and  an equivalent bilayer graphene. The MSD can give a clear interpretation about the moir\'{e} band structure of the (2+1+2)-DTMLG. For example, since the $H_+$ is equal to a TMBG, the (2+1+2)-DTMLG has a pair of  topological flat bands with nonzero valley Chern number, which is exactly the same as the case of TMBG, but appears at a larger magic angle. 
	
	When $Y > 1$, the DTMLG is mapped into two twisted multilayer graphene: a $(X+\frac{Y+1}{2})$-type (single twist, even parity) and a $(X+\frac{Y-1}{2})$-type (single twist, odd parity). Here, the moir\'{e} interlayer hopping does not have the $\sqrt{2}$ factor, so that the two subsystems both can give rise to a pair of flat bands at the magic angle $1.05^\circ$. In other words, we can get fourfold degenerate flat bands in this case. The  (1+3+1)-DTMLG is the simplest example for this case, in which $H_+$ is equal to a TMBG, and $H_-$ is equal to a TBG. According to the MSD, we can well understand the behaviors of  the foudfold degenerate flat bands here. The two flat bands from TMBG are of even parity and  have nonzero valley Chern number. Meanwhile, the two from the TBG is of odd parity and are topological trivial\cite{ma2021doubled}.  
	
	The MSD indicates that the perpendicular electric field $E_\perp$ will act as  hybridization terms between $H_+$ and $H_-$, since $E_\perp$ break the mirror symmetry.　In other words, $E_\perp$ here has a different way to couple with the DTMLG, distinct from the single twist moir\'{e} heterostructures.
	
	Most importantly, the MSD implies that all the novel correlated states observed in the twisted multilayer graphene systems~(e.g., TBG\cite{cao2018correlated,cao2018unconventional,lu2019superconductors,sharpe2019}, TMBG\cite{chen2021TMBG,xu2021tmbg,polshyn2020tmbg,he2021tmbg,yin2022tmbg}, twisted double-bilayer graphene\cite{cao2020tunable,Tunable583,doi:10.1021/acs.nanolett.9b05117,zhangguangyu2019,Samajdar_2021},    etc.)~as well as the multilayer graphene (e.g.~bilayer graphene\cite{zhou2022science,blg2022natphy,blg2022nature}) can also exist in the DTMLG system, since their Hamiltonians are exactly equivalent. It is quite like the case of ATTLG, which can be decoupled into a TBG and a MLG. Since superconducting state is observed in TBG, it is  expected that the superconducting state should also exist in ATTLG, which is finally confirmed in recent experiments. Thus, according to the MSD above, we expect that, similar as the TBG case, the superconducting state should exist in the (1+3+1)-DTMLG. Meanwhile, due to the mirror symmetry, the parity (even or odd) becomes a new discrete degree of freedom of the electrons in the DTMLG, in addition to the spin and valley. Thus, in principle, we may also expect that some new correlated states, which spontaneously breaks the parity degree of freedom,  can exist in the DTMLG systems.  One possible system is the (1+3+1)-DTFLG, where the parity polarized states may exist in the fourfold degenerate flat bands, when the Coulomb interaction is considered. 
	
	{\em Note added.} In this paper, our decomposition method depends on the mirror symmetry and gives an exact mapping. We note that a chiral decomposition based on perturbation calculation is proposed  in recent works\cite{zhang2020chiral,  xie2022alternating}, which is distinct from our method.

	\begin{acknowledgments}
		This work was supported by the National Natural Science Foundation of China(Grants No.~12141401, No.~11874160, No.~22273029 and No.~11534001), the National Key Research and Development Program of China (No.~2017YFA040351), and the Fundamental Research Funds for the Central Universities(HUST:No. 2017KFYXJJ027).
	\end{acknowledgments}
	
	\begin{appendix}
		\section{Derivation of the mirror symmetry decomposition of a general DTMLG}
		Here we derive the decoupled Hamiltonian in Eq.~\eqref{Hmirr-general} and Eq.~\eqref{H+}. We start from the Hamiltonian of a general (X+Y+Z)-DTMLG with mirror symmetry as given Eq.~\eqref{Hgeneral}.
		
		Considering the mirror symmetry, we use the  parity resolved basis, as given in Eq.~\eqref{basis-mirror-general}.
		The corresponding unitary transformation is
		\begin{equation}\label{U-general}
			U=\frac{1}{\sqrt{2}}
			\left(
			\begin{array}{ccc}
				E& 0 &\gamma \\
				0& \sqrt{2}I& 0\\
				\gamma& 0&-E
			\end{array}
			\right).
		\end{equation}
		Here $E$ is a $2(N-1)\times 2(N-1)$ identity matrix, and $\gamma$ is a $2(N-1)\times 2(N-1)$ matrix as 
		\begin{equation}
			\gamma=\left(
			\begin{array}{ccc}
				0&\cdots&I\\
				\vdots&I&\vdots\\
				I&\cdots&0
			\end{array}
			\right).
		\end{equation}
		So the general Hamiltonian after this unitary transformtion is
		\begin{widetext}
			\begin{align}\label{H-general-help}
				H_{mirr}&=U^{-1}H_{0}U=\frac{1}{2}\left(
				\begin{array}{ccc}
					H_{down}+\gamma H_{up}\gamma&\sqrt{2}(T^\dag_{dm}+\gamma T_{um})&H_{down}\gamma-\gamma H_{up}\\
					\sqrt{2}(T_{dm}+T^\dag_{um}\gamma)&2H_N&\sqrt{2}(T_{dm}\gamma-T^\dag_{um})\\
					\gamma H_{down}-H_{up}\gamma&\sqrt{2}(\gamma T^\dag_{dm}-T_{um})&\gamma H_{down}\gamma +H_{up}
				\end{array}
				\right).
			\end{align}
		\end{widetext}
		
		Now we need to use the mirror symmetry, which implies
		\begin{equation}
			\begin{aligned}
				H&_1=H_L\\
				H&_2=H_{L-1}\\
				\qquad\vdots\\
				H&_{N-1}=H_{N+1}\\
				T&_{1,2}=T^{\dag}_{L-1,L}\\
				T&_{2,3}=T^{\dag}_{L-2,1}\\
				\qquad\vdots\\
				T&_{N-1,N}=T^{\dag}_{N,N+1}.
			\end{aligned}
		\end{equation}
		With the help of mirror symmetry, we also can verify $H_{down}=\gamma H_{up}\gamma$ and $T^{\dag}_{dm}=\gamma T_{um}$. From the  Eq.~\eqref{H-general-help}, we finally get 
		\begin{equation}
			H_{mirr}=\left(
			\begin{array}{ccc}
				H_{down}&\sqrt{2}T^\dag_{dm}&0\\
				\sqrt{2}T_{dm}&H_N&0\\
				0&0&H_{up}
			\end{array}
			\right),
		\end{equation}
		which is just the Eq.~\eqref{Hmirr-general} in the main text.

		\section{Derivation of the $H_V$}
		Here we give a derivation of the influence of perpendicular electric field $E_{\perp}$ under parity resolved basis, i.e.~the Eq.~\eqref{H-general-V} and Eq.~\eqref{V-general} in the main text. With the original Bloch wave basis \{$\ket{A_i}$, $\ket{B_i}$ \}, the Hamiltonian of $E_{\perp}$ is expressed as a block diagonal matrix
		\begin{align}
			&H_0^V=\notag\\
			&VI\otimes\left(
			\begin{array}{ccccccc}
				N-1&&&&&&\\
				&N-2&&&&&\\
				&&\ddots&&&&\\
				&&&0&&&\\
				&&&&\ddots&&\\
				&&&&&-(N-2)&\\
				&&&&&&-(N-1)
			\end{array}
			\right),
		\end{align}
		where V is the potential difference between adjacent layers. For convenience,  we define a new diagonal block matrix
		\begin{equation}
			\chi=I\otimes\left(
			\begin{array}{ccccc}
				N-1&&&&\\
				&N-2&&&\\
				&&\cdots&&\\
				&&&2&\\
				&&&&1
			\end{array}
			\right),
		\end{equation}
		 and  then
		\begin{equation}
			H_0^V=V\left(
			\begin{array}{ccc}
				\chi&0&0\\
				0&0&0\\
				0&0&-\gamma \chi\gamma
			\end{array}
			\right).
		\end{equation}
	
 In the parity resolved basis,  we have  
		\begin{align}
			H_V&=U^{-1}H_0^VU=V\left(
			\begin{array}{ccc}
				0&0&\chi\gamma\\
				0&0&0\\
				\gamma\chi&0&0
			\end{array}
			\right),
		\end{align}
where the unitary matrix is given  by Eq.~\eqref{U-general}.  
		We then define $\Delta=V\chi\gamma$,   and finally we get
		\begin{equation}
			H_V=\left(
			\begin{array}{cc}
				0&D\\
				D^\dag&0
			\end{array}
			\right),
		\end{equation}
  where D is given in Eq.\eqref{V-general} of the main text.
	\end{appendix}
	
	\bibliography{references}

\begin{thebibliography}{83}%
\makeatletter
\providecommand \@ifxundefined [1]{%
 \@ifx{#1\undefined}
}%
\providecommand \@ifnum [1]{%
 \ifnum #1\expandafter \@firstoftwo
 \else \expandafter \@secondoftwo
 \fi
}%
\providecommand \@ifx [1]{%
 \ifx #1\expandafter \@firstoftwo
 \else \expandafter \@secondoftwo
 \fi
}%
\providecommand \natexlab [1]{#1}%
\providecommand \enquote  [1]{``#1''}%
\providecommand \bibnamefont  [1]{#1}%
\providecommand \bibfnamefont [1]{#1}%
\providecommand \citenamefont [1]{#1}%
\providecommand \href@noop [0]{\@secondoftwo}%
\providecommand \href [0]{\begingroup \@sanitize@url \@href}%
\providecommand \@href[1]{\@@startlink{#1}\@@href}%
\providecommand \@@href[1]{\endgroup#1\@@endlink}%
\providecommand \@sanitize@url [0]{\catcode `\\12\catcode `\$12\catcode
  `\&12\catcode `\#12\catcode `\^12\catcode `\_12\catcode `\%12\relax}%
\providecommand \@@startlink[1]{}%
\providecommand \@@endlink[0]{}%
\providecommand \url  [0]{\begingroup\@sanitize@url \@url }%
\providecommand \@url [1]{\endgroup\@href {#1}{\urlprefix }}%
\providecommand \urlprefix  [0]{URL }%
\providecommand \Eprint [0]{\href }%
\providecommand \doibase [0]{http://dx.doi.org/}%
\providecommand \selectlanguage [0]{\@gobble}%
\providecommand \bibinfo  [0]{\@secondoftwo}%
\providecommand \bibfield  [0]{\@secondoftwo}%
\providecommand \translation [1]{[#1]}%
\providecommand \BibitemOpen [0]{}%
\providecommand \bibitemStop [0]{}%
\providecommand \bibitemNoStop [0]{.\EOS\space}%
\providecommand \EOS [0]{\spacefactor3000\relax}%
\providecommand \BibitemShut  [1]{\csname bibitem#1\endcsname}%
\let\auto@bib@innerbib\@empty
\bibitem [{\citenamefont {Park}\ \emph {et~al.}(2021)\citenamefont {Park},
  \citenamefont {Cao}, \citenamefont {Watanabe}, \citenamefont {Taniguchi},\
  and\ \citenamefont {Jarillo-Herrero}}]{park2021tunable}%
  \BibitemOpen
  \bibfield  {author} {\bibinfo {author} {\bibfnamefont {J.~M.}\ \bibnamefont
  {Park}}, \bibinfo {author} {\bibfnamefont {Y.}~\bibnamefont {Cao}}, \bibinfo
  {author} {\bibfnamefont {K.}~\bibnamefont {Watanabe}}, \bibinfo {author}
  {\bibfnamefont {T.}~\bibnamefont {Taniguchi}}, \ and\ \bibinfo {author}
  {\bibfnamefont {P.}~\bibnamefont {Jarillo-Herrero}},\ }\href {\doibase
  10.1038/s41586-021-03192-0} {\bibfield  {journal} {\bibinfo  {journal}
  {Nature}\ }\textbf {\bibinfo {volume} {590}},\ \bibinfo {pages} {249}
  (\bibinfo {year} {2021})}\BibitemShut {NoStop}%
\bibitem [{\citenamefont {Hao}\ \emph {et~al.}(2021)\citenamefont {Hao},
  \citenamefont {Zimmerman}, \citenamefont {Ledwith}, \citenamefont {Khalaf},
  \citenamefont {Najafabadi}, \citenamefont {Watanabe}, \citenamefont
  {Taniguchi}, \citenamefont {Vishwanath},\ and\ \citenamefont
  {Kim}}]{Haoscience2021}%
  \BibitemOpen
  \bibfield  {author} {\bibinfo {author} {\bibfnamefont {Z.}~\bibnamefont
  {Hao}}, \bibinfo {author} {\bibfnamefont {A.~M.}\ \bibnamefont {Zimmerman}},
  \bibinfo {author} {\bibfnamefont {P.}~\bibnamefont {Ledwith}}, \bibinfo
  {author} {\bibfnamefont {E.}~\bibnamefont {Khalaf}}, \bibinfo {author}
  {\bibfnamefont {D.~H.}\ \bibnamefont {Najafabadi}}, \bibinfo {author}
  {\bibfnamefont {K.}~\bibnamefont {Watanabe}}, \bibinfo {author}
  {\bibfnamefont {T.}~\bibnamefont {Taniguchi}}, \bibinfo {author}
  {\bibfnamefont {A.}~\bibnamefont {Vishwanath}}, \ and\ \bibinfo {author}
  {\bibfnamefont {P.}~\bibnamefont {Kim}},\ }\href {\doibase
  10.1126/science.abg0399} {\bibfield  {journal} {\bibinfo  {journal}
  {Science}\ }\textbf {\bibinfo {volume} {371}},\ \bibinfo {pages} {1133}
  (\bibinfo {year} {2021})}\BibitemShut {NoStop}%
\bibitem [{\citenamefont {Cao}\ \emph {et~al.}(2021{\natexlab{a}})\citenamefont
  {Cao}, \citenamefont {Park}, \citenamefont {Watanabe}, \citenamefont
  {Taniguchi},\ and\ \citenamefont {Jarillo-Herrero}}]{cao2021pauli}%
  \BibitemOpen
  \bibfield  {author} {\bibinfo {author} {\bibfnamefont {Y.}~\bibnamefont
  {Cao}}, \bibinfo {author} {\bibfnamefont {J.~M.}\ \bibnamefont {Park}},
  \bibinfo {author} {\bibfnamefont {K.}~\bibnamefont {Watanabe}}, \bibinfo
  {author} {\bibfnamefont {T.}~\bibnamefont {Taniguchi}}, \ and\ \bibinfo
  {author} {\bibfnamefont {P.}~\bibnamefont {Jarillo-Herrero}},\ }\href
  {\doibase 10.1038/s41586-021-03685-y} {\bibfield  {journal} {\bibinfo
  {journal} {Nature}\ }\textbf {\bibinfo {volume} {595}},\ \bibinfo {pages}
  {526} (\bibinfo {year} {2021}{\natexlab{a}})}\BibitemShut {NoStop}%
\bibitem [{\citenamefont {Kim}\ \emph {et~al.}(2022)\citenamefont {Kim},
  \citenamefont {Choi}, \citenamefont {Lewandowski}, \citenamefont {Thomson},
  \citenamefont {Zhang}, \citenamefont {Polski}, \citenamefont {Watanabe},
  \citenamefont {Taniguchi}, \citenamefont {Alicea},\ and\ \citenamefont
  {Nadj-Perge}}]{nature2022}%
  \BibitemOpen
  \bibfield  {author} {\bibinfo {author} {\bibfnamefont {H.}~\bibnamefont
  {Kim}}, \bibinfo {author} {\bibfnamefont {Y.}~\bibnamefont {Choi}}, \bibinfo
  {author} {\bibfnamefont {C.}~\bibnamefont {Lewandowski}}, \bibinfo {author}
  {\bibfnamefont {A.}~\bibnamefont {Thomson}}, \bibinfo {author} {\bibfnamefont
  {Y.}~\bibnamefont {Zhang}}, \bibinfo {author} {\bibfnamefont
  {R.}~\bibnamefont {Polski}}, \bibinfo {author} {\bibfnamefont
  {K.}~\bibnamefont {Watanabe}}, \bibinfo {author} {\bibfnamefont
  {T.}~\bibnamefont {Taniguchi}}, \bibinfo {author} {\bibfnamefont
  {J.}~\bibnamefont {Alicea}}, \ and\ \bibinfo {author} {\bibfnamefont
  {S.}~\bibnamefont {Nadj-Perge}},\ }\href {\doibase
  10.1038/s41586-022-04715-z} {\bibfield  {journal} {\bibinfo  {journal}
  {Nature}\ }\textbf {\bibinfo {volume} {606}},\ \bibinfo {pages} {494}
  (\bibinfo {year} {2022})}\BibitemShut {NoStop}%
\bibitem [{\citenamefont {{Liu}}\ \emph {et~al.}(2022)\citenamefont {{Liu}},
  \citenamefont {{Zhang}}, \citenamefont {{Watanabe}}, \citenamefont
  {{Taniguchi}},\ and\ \citenamefont {{Li}}}]{liuxiaoxue2022natphy}%
  \BibitemOpen
  \bibfield  {author} {\bibinfo {author} {\bibfnamefont {X.}~\bibnamefont
  {{Liu}}}, \bibinfo {author} {\bibfnamefont {N.~J.}\ \bibnamefont {{Zhang}}},
  \bibinfo {author} {\bibfnamefont {K.}~\bibnamefont {{Watanabe}}}, \bibinfo
  {author} {\bibfnamefont {T.}~\bibnamefont {{Taniguchi}}}, \ and\ \bibinfo
  {author} {\bibfnamefont {J.~I.~A.}\ \bibnamefont {{Li}}},\ }\href {\doibase
  10.1038/s41567-022-01515-0} {\bibfield  {journal} {\bibinfo  {journal} {Nat.
  Phys.}\ }\textbf {\bibinfo {volume} {18}},\ \bibinfo {pages} {522} (\bibinfo
  {year} {2022})}\BibitemShut {NoStop}%
\bibitem [{\citenamefont {Turkel}\ \emph {et~al.}(2022)\citenamefont {Turkel},
  \citenamefont {Swann}, \citenamefont {Zhu}, \citenamefont {Christos},
  \citenamefont {Watanabe}, \citenamefont {Taniguchi}, \citenamefont {Sachdev},
  \citenamefont {Scheurer}, \citenamefont {Kaxiras}, \citenamefont {Dean},\
  and\ \citenamefont {Pasupathy}}]{science2022}%
  \BibitemOpen
  \bibfield  {author} {\bibinfo {author} {\bibfnamefont {S.}~\bibnamefont
  {Turkel}}, \bibinfo {author} {\bibfnamefont {J.}~\bibnamefont {Swann}},
  \bibinfo {author} {\bibfnamefont {Z.}~\bibnamefont {Zhu}}, \bibinfo {author}
  {\bibfnamefont {M.}~\bibnamefont {Christos}}, \bibinfo {author}
  {\bibfnamefont {K.}~\bibnamefont {Watanabe}}, \bibinfo {author}
  {\bibfnamefont {T.}~\bibnamefont {Taniguchi}}, \bibinfo {author}
  {\bibfnamefont {S.}~\bibnamefont {Sachdev}}, \bibinfo {author} {\bibfnamefont
  {M.}~\bibnamefont {Scheurer}}, \bibinfo {author} {\bibfnamefont
  {E.}~\bibnamefont {Kaxiras}}, \bibinfo {author} {\bibfnamefont
  {C.}~\bibnamefont {Dean}}, \ and\ \bibinfo {author} {\bibfnamefont
  {A.}~\bibnamefont {Pasupathy}},\ }\href {\doibase 10.1126/science.abk1895}
  {\bibfield  {journal} {\bibinfo  {journal} {Science}\ }\textbf {\bibinfo
  {volume} {376}},\ \bibinfo {pages} {193} (\bibinfo {year}
  {2022})}\BibitemShut {NoStop}%
\bibitem [{\citenamefont {Khalaf}\ \emph {et~al.}(2019)\citenamefont {Khalaf},
  \citenamefont {Kruchkov}, \citenamefont {Tarnopolsky},\ and\ \citenamefont
  {Vishwanath}}]{Khalaf2019prb}%
  \BibitemOpen
  \bibfield  {author} {\bibinfo {author} {\bibfnamefont {E.}~\bibnamefont
  {Khalaf}}, \bibinfo {author} {\bibfnamefont {A.~J.}\ \bibnamefont
  {Kruchkov}}, \bibinfo {author} {\bibfnamefont {G.}~\bibnamefont
  {Tarnopolsky}}, \ and\ \bibinfo {author} {\bibfnamefont {A.}~\bibnamefont
  {Vishwanath}},\ }\href {\doibase 10.1103/PhysRevB.100.085109} {\bibfield
  {journal} {\bibinfo  {journal} {Phys. Rev. B}\ }\textbf {\bibinfo {volume}
  {100}},\ \bibinfo {pages} {085109} (\bibinfo {year} {2019})}\BibitemShut
  {NoStop}%
\bibitem [{\citenamefont {Carr}\ \emph {et~al.}(2020)\citenamefont {Carr},
  \citenamefont {Li}, \citenamefont {Zhu}, \citenamefont {Kaxiras},
  \citenamefont {Sachdev},\ and\ \citenamefont {Kruchkov}}]{carr2020nanolett}%
  \BibitemOpen
  \bibfield  {author} {\bibinfo {author} {\bibfnamefont {S.}~\bibnamefont
  {Carr}}, \bibinfo {author} {\bibfnamefont {C.}~\bibnamefont {Li}}, \bibinfo
  {author} {\bibfnamefont {Z.}~\bibnamefont {Zhu}}, \bibinfo {author}
  {\bibfnamefont {E.}~\bibnamefont {Kaxiras}}, \bibinfo {author} {\bibfnamefont
  {S.}~\bibnamefont {Sachdev}}, \ and\ \bibinfo {author} {\bibfnamefont
  {A.}~\bibnamefont {Kruchkov}},\ }\href {\doibase
  10.1021/acs.nanolett.9b04979} {\bibfield  {journal} {\bibinfo  {journal}
  {Nano. Lett.}\ }\textbf {\bibinfo {volume} {20}},\ \bibinfo {pages} {3030}
  (\bibinfo {year} {2020})}\BibitemShut {NoStop}%
\bibitem [{\citenamefont {{Li}}\ \emph {et~al.}()\citenamefont {{Li}},
  \citenamefont {{Wu}},\ and\ \citenamefont {{MacDonald}}}]{lixiao2019arXiv}%
  \BibitemOpen
  \bibfield  {author} {\bibinfo {author} {\bibfnamefont {X.}~\bibnamefont
  {{Li}}}, \bibinfo {author} {\bibfnamefont {F.}~\bibnamefont {{Wu}}}, \ and\
  \bibinfo {author} {\bibfnamefont {A.~H.}\ \bibnamefont {{MacDonald}}},\
  }\href@noop {} {\ }\Eprint {http://arxiv.org/abs/1907.12338}
  {arXiv:1907.12338} \BibitemShut {NoStop}%
\bibitem [{\citenamefont {Lei}\ \emph {et~al.}(2021)\citenamefont {Lei},
  \citenamefont {Linhart}, \citenamefont {Qin}, \citenamefont {Libisch},\ and\
  \citenamefont {MacDonald}}]{Leichao2021PRB}%
  \BibitemOpen
  \bibfield  {author} {\bibinfo {author} {\bibfnamefont {C.}~\bibnamefont
  {Lei}}, \bibinfo {author} {\bibfnamefont {L.}~\bibnamefont {Linhart}},
  \bibinfo {author} {\bibfnamefont {W.}~\bibnamefont {Qin}}, \bibinfo {author}
  {\bibfnamefont {F.}~\bibnamefont {Libisch}}, \ and\ \bibinfo {author}
  {\bibfnamefont {A.~H.}\ \bibnamefont {MacDonald}},\ }\href {\doibase
  10.1103/PhysRevB.104.035139} {\bibfield  {journal} {\bibinfo  {journal}
  {Phys. Rev. B}\ }\textbf {\bibinfo {volume} {104}},\ \bibinfo {pages}
  {035139} (\bibinfo {year} {2021})}\BibitemShut {NoStop}%
\bibitem [{\citenamefont {Wu}\ \emph {et~al.}(2021)\citenamefont {Wu},
  \citenamefont {Zhan},\ and\ \citenamefont {Yuan}}]{yuan2021china}%
  \BibitemOpen
  \bibfield  {author} {\bibinfo {author} {\bibfnamefont {Z.}~\bibnamefont
  {Wu}}, \bibinfo {author} {\bibfnamefont {Z.}~\bibnamefont {Zhan}}, \ and\
  \bibinfo {author} {\bibfnamefont {S.}~\bibnamefont {Yuan}},\ }\href {\doibase
  10.1007/s11433-020-1690-4} {\bibfield  {journal} {\bibinfo  {journal} {Sci.
  China Phys. Mech. Astron.}\ }\textbf {\bibinfo {volume} {64}},\ \bibinfo
  {pages} {1} (\bibinfo {year} {2021})}\BibitemShut {NoStop}%
\bibitem [{\citenamefont {Ledwith}\ \emph {et~al.}(2021)\citenamefont
  {Ledwith}, \citenamefont {Khalaf}, \citenamefont {Zhu}, \citenamefont {Carr},
  \citenamefont {Kaxiras},\ and\ \citenamefont {Vishwanath}}]{ledwith2021tb}%
  \BibitemOpen
  \bibfield  {author} {\bibinfo {author} {\bibfnamefont {P.~J.}\ \bibnamefont
  {Ledwith}}, \bibinfo {author} {\bibfnamefont {E.}~\bibnamefont {Khalaf}},
  \bibinfo {author} {\bibfnamefont {Z.}~\bibnamefont {Zhu}}, \bibinfo {author}
  {\bibfnamefont {S.}~\bibnamefont {Carr}}, \bibinfo {author} {\bibfnamefont
  {E.}~\bibnamefont {Kaxiras}}, \ and\ \bibinfo {author} {\bibfnamefont
  {A.}~\bibnamefont {Vishwanath}},\ }\href@noop {} {\  (\bibinfo {year}
  {2021})},\ \Eprint {http://arxiv.org/abs/2111.11060} {arXiv:2111.11060}
  \BibitemShut {NoStop}%
\bibitem [{\citenamefont {Shin}\ \emph {et~al.}(2021)\citenamefont {Shin},
  \citenamefont {Chittari},\ and\ \citenamefont {Jung}}]{shin2021prb}%
  \BibitemOpen
  \bibfield  {author} {\bibinfo {author} {\bibfnamefont {J.}~\bibnamefont
  {Shin}}, \bibinfo {author} {\bibfnamefont {B.~L.}\ \bibnamefont {Chittari}},
  \ and\ \bibinfo {author} {\bibfnamefont {J.}~\bibnamefont {Jung}},\ }\href
  {\doibase 10.1103/PhysRevB.104.045413} {\bibfield  {journal} {\bibinfo
  {journal} {Phys. Rev. B}\ }\textbf {\bibinfo {volume} {104}},\ \bibinfo
  {pages} {045413} (\bibinfo {year} {2021})}\BibitemShut {NoStop}%
\bibitem [{\citenamefont {C\ifmmode \u{a}\else \u{a}\fi{}lug\ifmmode~\u{a}\else
  \u{a}\fi{}ru}\ \emph {et~al.}(2021)\citenamefont {C\ifmmode \u{a}\else
  \u{a}\fi{}lug\ifmmode~\u{a}\else \u{a}\fi{}ru}, \citenamefont {Xie},
  \citenamefont {Song}, \citenamefont {Lian}, \citenamefont {Regnault},\ and\
  \citenamefont {Bernevig}}]{xiefang2021prb}%
  \BibitemOpen
  \bibfield  {author} {\bibinfo {author} {\bibfnamefont {D.}~\bibnamefont
  {C\ifmmode \u{a}\else \u{a}\fi{}lug\ifmmode~\u{a}\else \u{a}\fi{}ru}},
  \bibinfo {author} {\bibfnamefont {F.}~\bibnamefont {Xie}}, \bibinfo {author}
  {\bibfnamefont {Z.-D.}\ \bibnamefont {Song}}, \bibinfo {author}
  {\bibfnamefont {B.}~\bibnamefont {Lian}}, \bibinfo {author} {\bibfnamefont
  {N.}~\bibnamefont {Regnault}}, \ and\ \bibinfo {author} {\bibfnamefont
  {B.~A.}\ \bibnamefont {Bernevig}},\ }\href {\doibase
  10.1103/PhysRevB.103.195411} {\bibfield  {journal} {\bibinfo  {journal}
  {Phys. Rev. B}\ }\textbf {\bibinfo {volume} {103}},\ \bibinfo {pages}
  {195411} (\bibinfo {year} {2021})}\BibitemShut {NoStop}%
\bibitem [{\citenamefont {Xie}\ \emph {et~al.}(2021)\citenamefont {Xie},
  \citenamefont {Regnault}, \citenamefont {C\ifmmode \u{a}\else
  \u{a}\fi{}lug\ifmmode~\u{a}\else \u{a}\fi{}ru}, \citenamefont {Bernevig},\
  and\ \citenamefont {Lian}}]{lianbiao2021prb}%
  \BibitemOpen
  \bibfield  {author} {\bibinfo {author} {\bibfnamefont {F.}~\bibnamefont
  {Xie}}, \bibinfo {author} {\bibfnamefont {N.}~\bibnamefont {Regnault}},
  \bibinfo {author} {\bibfnamefont {D.}~\bibnamefont {C\ifmmode \u{a}\else
  \u{a}\fi{}lug\ifmmode~\u{a}\else \u{a}\fi{}ru}}, \bibinfo {author}
  {\bibfnamefont {B.~A.}\ \bibnamefont {Bernevig}}, \ and\ \bibinfo {author}
  {\bibfnamefont {B.}~\bibnamefont {Lian}},\ }\href {\doibase
  10.1103/PhysRevB.104.115167} {\bibfield  {journal} {\bibinfo  {journal}
  {Phys. Rev. B}\ }\textbf {\bibinfo {volume} {104}},\ \bibinfo {pages}
  {115167} (\bibinfo {year} {2021})}\BibitemShut {NoStop}%
\bibitem [{\citenamefont {Ramires}\ and\ \citenamefont
  {Lado}(2021)}]{lado2021prl}%
  \BibitemOpen
  \bibfield  {author} {\bibinfo {author} {\bibfnamefont {A.}~\bibnamefont
  {Ramires}}\ and\ \bibinfo {author} {\bibfnamefont {J.~L.}\ \bibnamefont
  {Lado}},\ }\href {\doibase 10.1103/PhysRevLett.127.026401} {\bibfield
  {journal} {\bibinfo  {journal} {Phys. Rev. Lett.}\ }\textbf {\bibinfo
  {volume} {127}},\ \bibinfo {pages} {026401} (\bibinfo {year}
  {2021})}\BibitemShut {NoStop}%
\bibitem [{\citenamefont {Christos}\ \emph {et~al.}(2022)\citenamefont
  {Christos}, \citenamefont {Sachdev},\ and\ \citenamefont
  {Scheurer}}]{prx2022}%
  \BibitemOpen
  \bibfield  {author} {\bibinfo {author} {\bibfnamefont {M.}~\bibnamefont
  {Christos}}, \bibinfo {author} {\bibfnamefont {S.}~\bibnamefont {Sachdev}}, \
  and\ \bibinfo {author} {\bibfnamefont {M.~S.}\ \bibnamefont {Scheurer}},\
  }\href {\doibase 10.1103/PhysRevX.12.021018} {\bibfield  {journal} {\bibinfo
  {journal} {Phys. Rev. X}\ }\textbf {\bibinfo {volume} {12}},\ \bibinfo
  {pages} {021018} (\bibinfo {year} {2022})}\BibitemShut {NoStop}%
\bibitem [{\citenamefont {Lin}\ \emph {et~al.}(2022)\citenamefont {Lin},
  \citenamefont {Li}, \citenamefont {Su},\ and\ \citenamefont
  {Ni}}]{nijun2022prb}%
  \BibitemOpen
  \bibfield  {author} {\bibinfo {author} {\bibfnamefont {X.}~\bibnamefont
  {Lin}}, \bibinfo {author} {\bibfnamefont {C.}~\bibnamefont {Li}}, \bibinfo
  {author} {\bibfnamefont {K.}~\bibnamefont {Su}}, \ and\ \bibinfo {author}
  {\bibfnamefont {J.}~\bibnamefont {Ni}},\ }\href {\doibase
  10.1103/PhysRevB.106.075423} {\bibfield  {journal} {\bibinfo  {journal}
  {Phys. Rev. B}\ }\textbf {\bibinfo {volume} {106}},\ \bibinfo {pages}
  {075423} (\bibinfo {year} {2022})}\BibitemShut {NoStop}%
\bibitem [{\citenamefont {Lake}\ and\ \citenamefont
  {Senthil}(2021)}]{senthil2021prb}%
  \BibitemOpen
  \bibfield  {author} {\bibinfo {author} {\bibfnamefont {E.}~\bibnamefont
  {Lake}}\ and\ \bibinfo {author} {\bibfnamefont {T.}~\bibnamefont {Senthil}},\
  }\href {\doibase 10.1103/PhysRevB.104.174505} {\bibfield  {journal} {\bibinfo
   {journal} {Phys. Rev. B}\ }\textbf {\bibinfo {volume} {104}},\ \bibinfo
  {pages} {174505} (\bibinfo {year} {2021})}\BibitemShut {NoStop}%
\bibitem [{\citenamefont {Chou}\ \emph {et~al.}(2021)\citenamefont {Chou},
  \citenamefont {Wu}, \citenamefont {Sau},\ and\ \citenamefont
  {Das~Sarma}}]{sarma2021prl}%
  \BibitemOpen
  \bibfield  {author} {\bibinfo {author} {\bibfnamefont {Y.-Z.}\ \bibnamefont
  {Chou}}, \bibinfo {author} {\bibfnamefont {F.}~\bibnamefont {Wu}}, \bibinfo
  {author} {\bibfnamefont {J.~D.}\ \bibnamefont {Sau}}, \ and\ \bibinfo
  {author} {\bibfnamefont {S.}~\bibnamefont {Das~Sarma}},\ }\href {\doibase
  10.1103/PhysRevLett.127.217001} {\bibfield  {journal} {\bibinfo  {journal}
  {Phys. Rev. Lett.}\ }\textbf {\bibinfo {volume} {127}},\ \bibinfo {pages}
  {217001} (\bibinfo {year} {2021})}\BibitemShut {NoStop}%
\bibitem [{\citenamefont {Qin}\ and\ \citenamefont
  {MacDonald}(2021)}]{qin2021prl}%
  \BibitemOpen
  \bibfield  {author} {\bibinfo {author} {\bibfnamefont {W.}~\bibnamefont
  {Qin}}\ and\ \bibinfo {author} {\bibfnamefont {A.~H.}\ \bibnamefont
  {MacDonald}},\ }\href {\doibase 10.1103/PhysRevLett.127.097001} {\bibfield
  {journal} {\bibinfo  {journal} {Phys. Rev. Lett.}\ }\textbf {\bibinfo
  {volume} {127}},\ \bibinfo {pages} {097001} (\bibinfo {year}
  {2021})}\BibitemShut {NoStop}%
\bibitem [{\citenamefont {Phong}\ \emph {et~al.}(2021)\citenamefont {Phong},
  \citenamefont {Pantale\'on}, \citenamefont {Cea},\ and\ \citenamefont
  {Guinea}}]{guinea2021prb}%
  \BibitemOpen
  \bibfield  {author} {\bibinfo {author} {\bibfnamefont {V.~o.~T.}\
  \bibnamefont {Phong}}, \bibinfo {author} {\bibfnamefont {P.~A.}\ \bibnamefont
  {Pantale\'on}}, \bibinfo {author} {\bibfnamefont {T.}~\bibnamefont {Cea}}, \
  and\ \bibinfo {author} {\bibfnamefont {F.}~\bibnamefont {Guinea}},\ }\href
  {\doibase 10.1103/PhysRevB.104.L121116} {\bibfield  {journal} {\bibinfo
  {journal} {Phys. Rev. B}\ }\textbf {\bibinfo {volume} {104}},\ \bibinfo
  {pages} {L121116} (\bibinfo {year} {2021})}\BibitemShut {NoStop}%
\bibitem [{\citenamefont {Assi}\ \emph {et~al.}(2021)\citenamefont {Assi},
  \citenamefont {LeBlanc}, \citenamefont {Rodriguez-Vega}, \citenamefont
  {Bahlouli},\ and\ \citenamefont {Vogl}}]{vogl2021prb}%
  \BibitemOpen
  \bibfield  {author} {\bibinfo {author} {\bibfnamefont {I.~A.}\ \bibnamefont
  {Assi}}, \bibinfo {author} {\bibfnamefont {J.~P.~F.}\ \bibnamefont
  {LeBlanc}}, \bibinfo {author} {\bibfnamefont {M.}~\bibnamefont
  {Rodriguez-Vega}}, \bibinfo {author} {\bibfnamefont {H.}~\bibnamefont
  {Bahlouli}}, \ and\ \bibinfo {author} {\bibfnamefont {M.}~\bibnamefont
  {Vogl}},\ }\href {\doibase 10.1103/PhysRevB.104.195429} {\bibfield  {journal}
  {\bibinfo  {journal} {Phys. Rev. B}\ }\textbf {\bibinfo {volume} {104}},\
  \bibinfo {pages} {195429} (\bibinfo {year} {2021})}\BibitemShut {NoStop}%
\bibitem [{\citenamefont {Scammell}\ \emph {et~al.}(2022)\citenamefont
  {Scammell}, \citenamefont {Li},\ and\ \citenamefont
  {Scheurer}}]{2Dmaterial2022}%
  \BibitemOpen
  \bibfield  {author} {\bibinfo {author} {\bibfnamefont {H.}~\bibnamefont
  {Scammell}}, \bibinfo {author} {\bibfnamefont {J.}~\bibnamefont {Li}}, \ and\
  \bibinfo {author} {\bibfnamefont {M.}~\bibnamefont {Scheurer}},\ }\href
  {\doibase 10.1088/2053-1583/ac5b16} {\bibfield  {journal} {\bibinfo
  {journal} {2D Mater.}\ }\textbf {\bibinfo {volume} {9}},\ \bibinfo {pages}
  {025027} (\bibinfo {year} {2022})}\BibitemShut {NoStop}%
\bibitem [{\citenamefont {Samajdar}\ \emph {et~al.}(2022)\citenamefont
  {Samajdar}, \citenamefont {Teng},\ and\ \citenamefont
  {Scheurer}}]{phonon2022prb}%
  \BibitemOpen
  \bibfield  {author} {\bibinfo {author} {\bibfnamefont {R.}~\bibnamefont
  {Samajdar}}, \bibinfo {author} {\bibfnamefont {Y.}~\bibnamefont {Teng}}, \
  and\ \bibinfo {author} {\bibfnamefont {M.~S.}\ \bibnamefont {Scheurer}},\
  }\href {\doibase 10.1103/PhysRevB.106.L201403} {\bibfield  {journal}
  {\bibinfo  {journal} {Phys. Rev. B}\ }\textbf {\bibinfo {volume} {106}},\
  \bibinfo {pages} {L201403} (\bibinfo {year} {2022})}\BibitemShut {NoStop}%
\bibitem [{\citenamefont {Guerci}\ \emph {et~al.}(2022)\citenamefont {Guerci},
  \citenamefont {Simon},\ and\ \citenamefont {Mora}}]{mora2022prr}%
  \BibitemOpen
  \bibfield  {author} {\bibinfo {author} {\bibfnamefont {D.}~\bibnamefont
  {Guerci}}, \bibinfo {author} {\bibfnamefont {P.}~\bibnamefont {Simon}}, \
  and\ \bibinfo {author} {\bibfnamefont {C.}~\bibnamefont {Mora}},\ }\href
  {\doibase 10.1103/PhysRevResearch.4.L012013} {\bibfield  {journal} {\bibinfo
  {journal} {Phys. Rev. Res.}\ }\textbf {\bibinfo {volume} {4}},\ \bibinfo
  {pages} {L012013} (\bibinfo {year} {2022})}\BibitemShut {NoStop}%
\bibitem [{\citenamefont {Li}\ \emph {et~al.}(2022)\citenamefont {Li},
  \citenamefont {Zhang}, \citenamefont {Chen}, \citenamefont {Wei},
  \citenamefont {Zhang}, \citenamefont {Xiao}, \citenamefont {Gao},
  \citenamefont {Chen}, \citenamefont {Liang}, \citenamefont {Pei},
  \citenamefont {Xu}, \citenamefont {Watanabe}, \citenamefont {Taniguchi},
  \citenamefont {Yang}, \citenamefont {Miao}, \citenamefont {Liu},
  \citenamefont {Cheng}, \citenamefont {Wang}, \citenamefont {Chen},\ and\
  \citenamefont {Liu}}]{adma2022}%
  \BibitemOpen
  \bibfield  {author} {\bibinfo {author} {\bibfnamefont {Y.}~\bibnamefont
  {Li}}, \bibinfo {author} {\bibfnamefont {S.}~\bibnamefont {Zhang}}, \bibinfo
  {author} {\bibfnamefont {F.}~\bibnamefont {Chen}}, \bibinfo {author}
  {\bibfnamefont {L.}~\bibnamefont {Wei}}, \bibinfo {author} {\bibfnamefont
  {Z.}~\bibnamefont {Zhang}}, \bibinfo {author} {\bibfnamefont
  {H.}~\bibnamefont {Xiao}}, \bibinfo {author} {\bibfnamefont {H.}~\bibnamefont
  {Gao}}, \bibinfo {author} {\bibfnamefont {M.}~\bibnamefont {Chen}}, \bibinfo
  {author} {\bibfnamefont {S.}~\bibnamefont {Liang}}, \bibinfo {author}
  {\bibfnamefont {D.}~\bibnamefont {Pei}}, \bibinfo {author} {\bibfnamefont
  {L.}~\bibnamefont {Xu}}, \bibinfo {author} {\bibfnamefont {K.}~\bibnamefont
  {Watanabe}}, \bibinfo {author} {\bibfnamefont {T.}~\bibnamefont {Taniguchi}},
  \bibinfo {author} {\bibfnamefont {L.}~\bibnamefont {Yang}}, \bibinfo {author}
  {\bibfnamefont {F.}~\bibnamefont {Miao}}, \bibinfo {author} {\bibfnamefont
  {J.}~\bibnamefont {Liu}}, \bibinfo {author} {\bibfnamefont {B.}~\bibnamefont
  {Cheng}}, \bibinfo {author} {\bibfnamefont {M.}~\bibnamefont {Wang}},
  \bibinfo {author} {\bibfnamefont {Y.}~\bibnamefont {Chen}}, \ and\ \bibinfo
  {author} {\bibfnamefont {Z.}~\bibnamefont {Liu}},\ }\href {\doibase
  https://doi.org/10.1002/adma.202205996} {\bibfield  {journal} {\bibinfo
  {journal} {Adv. Mater.}\ }\textbf {\bibinfo {volume} {34}},\ \bibinfo {pages}
  {2205996} (\bibinfo {year} {2022})}\BibitemShut {NoStop}%
\bibitem [{\citenamefont {Fischer}\ \emph {et~al.}(2022)\citenamefont
  {Fischer}, \citenamefont {Goodwin}, \citenamefont {Mostofi}, \citenamefont
  {Lischner}, \citenamefont {Kennes},\ and\ \citenamefont {Klebl}}]{npj2022}%
  \BibitemOpen
  \bibfield  {author} {\bibinfo {author} {\bibfnamefont {A.}~\bibnamefont
  {Fischer}}, \bibinfo {author} {\bibfnamefont {Z.}~\bibnamefont {Goodwin}},
  \bibinfo {author} {\bibfnamefont {A.}~\bibnamefont {Mostofi}}, \bibinfo
  {author} {\bibfnamefont {J.}~\bibnamefont {Lischner}}, \bibinfo {author}
  {\bibfnamefont {D.}~\bibnamefont {Kennes}}, \ and\ \bibinfo {author}
  {\bibfnamefont {L.}~\bibnamefont {Klebl}},\ }\href {\doibase
  10.1038/s41535-021-00410-w} {\bibfield  {journal} {\bibinfo  {journal} {npj
  Quant. Mater.}\ }\textbf {\bibinfo {volume} {7}},\ \bibinfo {pages} {5}
  (\bibinfo {year} {2022})}\BibitemShut {NoStop}%
\bibitem [{\citenamefont {Cao}\ \emph {et~al.}(2018{\natexlab{a}})\citenamefont
  {Cao}, \citenamefont {Fatemi}, \citenamefont {Demir}, \citenamefont {Fang},
  \citenamefont {Tomarken}, \citenamefont {Luo}, \citenamefont
  {Sanchez-Yamagishi}, \citenamefont {Watanabe}, \citenamefont {Taniguchi},
  \citenamefont {Kaxiras} \emph {et~al.}}]{cao2018correlated}%
  \BibitemOpen
  \bibfield  {author} {\bibinfo {author} {\bibfnamefont {Y.}~\bibnamefont
  {Cao}}, \bibinfo {author} {\bibfnamefont {V.}~\bibnamefont {Fatemi}},
  \bibinfo {author} {\bibfnamefont {A.}~\bibnamefont {Demir}}, \bibinfo
  {author} {\bibfnamefont {S.}~\bibnamefont {Fang}}, \bibinfo {author}
  {\bibfnamefont {S.~L.}\ \bibnamefont {Tomarken}}, \bibinfo {author}
  {\bibfnamefont {J.~Y.}\ \bibnamefont {Luo}}, \bibinfo {author} {\bibfnamefont
  {J.~D.}\ \bibnamefont {Sanchez-Yamagishi}}, \bibinfo {author} {\bibfnamefont
  {K.}~\bibnamefont {Watanabe}}, \bibinfo {author} {\bibfnamefont
  {T.}~\bibnamefont {Taniguchi}}, \bibinfo {author} {\bibfnamefont
  {E.}~\bibnamefont {Kaxiras}},  \emph {et~al.},\ }\href {\doibase
  10.1038/nature26154} {\bibfield  {journal} {\bibinfo  {journal} {Nature}\
  }\textbf {\bibinfo {volume} {556}},\ \bibinfo {pages} {80} (\bibinfo {year}
  {2018}{\natexlab{a}})}\BibitemShut {NoStop}%
\bibitem [{\citenamefont {Cao}\ \emph {et~al.}(2018{\natexlab{b}})\citenamefont
  {Cao}, \citenamefont {Fatemi}, \citenamefont {Fang}, \citenamefont
  {Watanabe}, \citenamefont {Taniguchi}, \citenamefont {Kaxiras},\ and\
  \citenamefont {Jarillo-Herrero}}]{cao2018unconventional}%
  \BibitemOpen
  \bibfield  {author} {\bibinfo {author} {\bibfnamefont {Y.}~\bibnamefont
  {Cao}}, \bibinfo {author} {\bibfnamefont {V.}~\bibnamefont {Fatemi}},
  \bibinfo {author} {\bibfnamefont {S.}~\bibnamefont {Fang}}, \bibinfo {author}
  {\bibfnamefont {K.}~\bibnamefont {Watanabe}}, \bibinfo {author}
  {\bibfnamefont {T.}~\bibnamefont {Taniguchi}}, \bibinfo {author}
  {\bibfnamefont {E.}~\bibnamefont {Kaxiras}}, \ and\ \bibinfo {author}
  {\bibfnamefont {P.}~\bibnamefont {Jarillo-Herrero}},\ }\href {\doibase
  10.1038/nature26160} {\bibfield  {journal} {\bibinfo  {journal} {Nature}\
  }\textbf {\bibinfo {volume} {556}},\ \bibinfo {pages} {43} (\bibinfo {year}
  {2018}{\natexlab{b}})}\BibitemShut {NoStop}%
\bibitem [{\citenamefont {Lu}\ \emph {et~al.}(2019)\citenamefont {Lu},
  \citenamefont {Stepanov}, \citenamefont {Yang}, \citenamefont {Xie},
  \citenamefont {Aamir}, \citenamefont {Das}, \citenamefont {Urgell},
  \citenamefont {Watanabe}, \citenamefont {Taniguchi}, \citenamefont {Zhang}
  \emph {et~al.}}]{lu2019superconductors}%
  \BibitemOpen
  \bibfield  {author} {\bibinfo {author} {\bibfnamefont {X.}~\bibnamefont
  {Lu}}, \bibinfo {author} {\bibfnamefont {P.}~\bibnamefont {Stepanov}},
  \bibinfo {author} {\bibfnamefont {W.}~\bibnamefont {Yang}}, \bibinfo {author}
  {\bibfnamefont {M.}~\bibnamefont {Xie}}, \bibinfo {author} {\bibfnamefont
  {M.~A.}\ \bibnamefont {Aamir}}, \bibinfo {author} {\bibfnamefont
  {I.}~\bibnamefont {Das}}, \bibinfo {author} {\bibfnamefont {C.}~\bibnamefont
  {Urgell}}, \bibinfo {author} {\bibfnamefont {K.}~\bibnamefont {Watanabe}},
  \bibinfo {author} {\bibfnamefont {T.}~\bibnamefont {Taniguchi}}, \bibinfo
  {author} {\bibfnamefont {G.}~\bibnamefont {Zhang}},  \emph {et~al.},\ }\href
  {\doibase 10.1038/s41586-019-1695-0} {\bibfield  {journal} {\bibinfo
  {journal} {Nature}\ }\textbf {\bibinfo {volume} {574}},\ \bibinfo {pages}
  {653} (\bibinfo {year} {2019})}\BibitemShut {NoStop}%
\bibitem [{\citenamefont {Cao}\ \emph {et~al.}(2020{\natexlab{a}})\citenamefont
  {Cao}, \citenamefont {Chowdhury}, \citenamefont {Rodan-Legrain},
  \citenamefont {Rubies-Bigorda}, \citenamefont {Watanabe}, \citenamefont
  {Taniguchi}, \citenamefont {Senthil},\ and\ \citenamefont
  {Jarillo-Herrero}}]{PhysRevLett.124.076801}%
  \BibitemOpen
  \bibfield  {author} {\bibinfo {author} {\bibfnamefont {Y.}~\bibnamefont
  {Cao}}, \bibinfo {author} {\bibfnamefont {D.}~\bibnamefont {Chowdhury}},
  \bibinfo {author} {\bibfnamefont {D.}~\bibnamefont {Rodan-Legrain}}, \bibinfo
  {author} {\bibfnamefont {O.}~\bibnamefont {Rubies-Bigorda}}, \bibinfo
  {author} {\bibfnamefont {K.}~\bibnamefont {Watanabe}}, \bibinfo {author}
  {\bibfnamefont {T.}~\bibnamefont {Taniguchi}}, \bibinfo {author}
  {\bibfnamefont {T.}~\bibnamefont {Senthil}}, \ and\ \bibinfo {author}
  {\bibfnamefont {P.}~\bibnamefont {Jarillo-Herrero}},\ }\href {\doibase
  10.1103/PhysRevLett.124.076801} {\bibfield  {journal} {\bibinfo  {journal}
  {Phys. Rev. Lett.}\ }\textbf {\bibinfo {volume} {124}},\ \bibinfo {pages}
  {076801} (\bibinfo {year} {2020}{\natexlab{a}})}\BibitemShut {NoStop}%
\bibitem [{\citenamefont {Andrei}\ and\ \citenamefont
  {MacDonald}(2020)}]{andrei2020graphene}%
  \BibitemOpen
  \bibfield  {author} {\bibinfo {author} {\bibfnamefont {E.~Y.}\ \bibnamefont
  {Andrei}}\ and\ \bibinfo {author} {\bibfnamefont {A.~H.}\ \bibnamefont
  {MacDonald}},\ }\href {\doibase 10.1038/s41563-020-00840-0} {\bibfield
  {journal} {\bibinfo  {journal} {Nat. Mater.}\ }\textbf {\bibinfo {volume}
  {19}},\ \bibinfo {pages} {1265} (\bibinfo {year} {2020})}\BibitemShut
  {NoStop}%
\bibitem [{\citenamefont {Polshyn}\ \emph {et~al.}(2019)\citenamefont
  {Polshyn}, \citenamefont {Yankowitz}, \citenamefont {Chen}, \citenamefont
  {Zhang}, \citenamefont {Watanabe}, \citenamefont {Taniguchi}, \citenamefont
  {Dean},\ and\ \citenamefont {Young}}]{polshyn2019}%
  \BibitemOpen
  \bibfield  {author} {\bibinfo {author} {\bibfnamefont {H.}~\bibnamefont
  {Polshyn}}, \bibinfo {author} {\bibfnamefont {M.}~\bibnamefont {Yankowitz}},
  \bibinfo {author} {\bibfnamefont {S.}~\bibnamefont {Chen}}, \bibinfo {author}
  {\bibfnamefont {Y.}~\bibnamefont {Zhang}}, \bibinfo {author} {\bibfnamefont
  {K.}~\bibnamefont {Watanabe}}, \bibinfo {author} {\bibfnamefont
  {T.}~\bibnamefont {Taniguchi}}, \bibinfo {author} {\bibfnamefont {C.~R.}\
  \bibnamefont {Dean}}, \ and\ \bibinfo {author} {\bibfnamefont {A.~F.}\
  \bibnamefont {Young}},\ }\href
  {https://www.nature.com/articles/s41567-019-0596-3} {\bibfield  {journal}
  {\bibinfo  {journal} {Nat. Phys.}\ }\textbf {\bibinfo {volume} {15}},\
  \bibinfo {pages} {1011} (\bibinfo {year} {2019})}\BibitemShut {NoStop}%
\bibitem [{\citenamefont {Lisi}\ \emph {et~al.}(2021)\citenamefont {Lisi},
  \citenamefont {Lu}, \citenamefont {Benschop}, \citenamefont {de~Jong},
  \citenamefont {Stepanov}, \citenamefont {Duran}, \citenamefont {Margot},
  \citenamefont {Cucchi}, \citenamefont {Cappelli}, \citenamefont {Hunter}
  \emph {et~al.}}]{lisi2021}%
  \BibitemOpen
  \bibfield  {author} {\bibinfo {author} {\bibfnamefont {S.}~\bibnamefont
  {Lisi}}, \bibinfo {author} {\bibfnamefont {X.}~\bibnamefont {Lu}}, \bibinfo
  {author} {\bibfnamefont {T.}~\bibnamefont {Benschop}}, \bibinfo {author}
  {\bibfnamefont {T.~A.}\ \bibnamefont {de~Jong}}, \bibinfo {author}
  {\bibfnamefont {P.}~\bibnamefont {Stepanov}}, \bibinfo {author}
  {\bibfnamefont {J.~R.}\ \bibnamefont {Duran}}, \bibinfo {author}
  {\bibfnamefont {F.}~\bibnamefont {Margot}}, \bibinfo {author} {\bibfnamefont
  {I.}~\bibnamefont {Cucchi}}, \bibinfo {author} {\bibfnamefont
  {E.}~\bibnamefont {Cappelli}}, \bibinfo {author} {\bibfnamefont
  {A.}~\bibnamefont {Hunter}},  \emph {et~al.},\ }\href
  {https://www.nature.com/articles/s41567-020-01041-x} {\bibfield  {journal}
  {\bibinfo  {journal} {Nat. Phys.}\ }\textbf {\bibinfo {volume} {17}},\
  \bibinfo {pages} {189} (\bibinfo {year} {2021})}\BibitemShut {NoStop}%
\bibitem [{\citenamefont {Choi}\ \emph {et~al.}(2019)\citenamefont {Choi},
  \citenamefont {Kemmer}, \citenamefont {Peng}, \citenamefont {Thomson},
  \citenamefont {Arora}, \citenamefont {Polski}, \citenamefont {Zhang},
  \citenamefont {Ren}, \citenamefont {Alicea}, \citenamefont {Refael} \emph
  {et~al.}}]{choi2019}%
  \BibitemOpen
  \bibfield  {author} {\bibinfo {author} {\bibfnamefont {Y.}~\bibnamefont
  {Choi}}, \bibinfo {author} {\bibfnamefont {J.}~\bibnamefont {Kemmer}},
  \bibinfo {author} {\bibfnamefont {Y.}~\bibnamefont {Peng}}, \bibinfo {author}
  {\bibfnamefont {A.}~\bibnamefont {Thomson}}, \bibinfo {author} {\bibfnamefont
  {H.}~\bibnamefont {Arora}}, \bibinfo {author} {\bibfnamefont
  {R.}~\bibnamefont {Polski}}, \bibinfo {author} {\bibfnamefont
  {Y.}~\bibnamefont {Zhang}}, \bibinfo {author} {\bibfnamefont
  {H.}~\bibnamefont {Ren}}, \bibinfo {author} {\bibfnamefont {J.}~\bibnamefont
  {Alicea}}, \bibinfo {author} {\bibfnamefont {G.}~\bibnamefont {Refael}},
  \emph {et~al.},\ }\href {https://www.nature.com/articles/s41567-019-0606-5}
  {\bibfield  {journal} {\bibinfo  {journal} {Nat. Phys.}\ }\textbf {\bibinfo
  {volume} {15}},\ \bibinfo {pages} {1174} (\bibinfo {year}
  {2019})}\BibitemShut {NoStop}%
\bibitem [{\citenamefont {Sharpe}\ \emph {et~al.}(2019)\citenamefont {Sharpe},
  \citenamefont {Fox}, \citenamefont {Barnard}, \citenamefont {Finney},
  \citenamefont {Watanabe}, \citenamefont {Taniguchi}, \citenamefont
  {Kastner},\ and\ \citenamefont {Goldhaber-Gordon}}]{sharpe2019}%
  \BibitemOpen
  \bibfield  {author} {\bibinfo {author} {\bibfnamefont {A.~L.}\ \bibnamefont
  {Sharpe}}, \bibinfo {author} {\bibfnamefont {E.~J.}\ \bibnamefont {Fox}},
  \bibinfo {author} {\bibfnamefont {A.~W.}\ \bibnamefont {Barnard}}, \bibinfo
  {author} {\bibfnamefont {J.}~\bibnamefont {Finney}}, \bibinfo {author}
  {\bibfnamefont {K.}~\bibnamefont {Watanabe}}, \bibinfo {author}
  {\bibfnamefont {T.}~\bibnamefont {Taniguchi}}, \bibinfo {author}
  {\bibfnamefont {M.}~\bibnamefont {Kastner}}, \ and\ \bibinfo {author}
  {\bibfnamefont {D.}~\bibnamefont {Goldhaber-Gordon}},\ }\href
  {https://www.science.org/doi/10.1126/science.aaw3780} {\bibfield  {journal}
  {\bibinfo  {journal} {Science}\ }\textbf {\bibinfo {volume} {365}},\ \bibinfo
  {pages} {605} (\bibinfo {year} {2019})}\BibitemShut {NoStop}%
\bibitem [{\citenamefont {Kazmierczak}\ \emph {et~al.}(2021)\citenamefont
  {Kazmierczak}, \citenamefont {Van~Winkle}, \citenamefont {Ophus},
  \citenamefont {Bustillo}, \citenamefont {Carr}, \citenamefont {Brown},
  \citenamefont {Ciston}, \citenamefont {Taniguchi}, \citenamefont {Watanabe},\
  and\ \citenamefont {Bediako}}]{kazmierczak2021strain}%
  \BibitemOpen
  \bibfield  {author} {\bibinfo {author} {\bibfnamefont {N.~P.}\ \bibnamefont
  {Kazmierczak}}, \bibinfo {author} {\bibfnamefont {M.}~\bibnamefont
  {Van~Winkle}}, \bibinfo {author} {\bibfnamefont {C.}~\bibnamefont {Ophus}},
  \bibinfo {author} {\bibfnamefont {K.~C.}\ \bibnamefont {Bustillo}}, \bibinfo
  {author} {\bibfnamefont {S.}~\bibnamefont {Carr}}, \bibinfo {author}
  {\bibfnamefont {H.~G.}\ \bibnamefont {Brown}}, \bibinfo {author}
  {\bibfnamefont {J.}~\bibnamefont {Ciston}}, \bibinfo {author} {\bibfnamefont
  {T.}~\bibnamefont {Taniguchi}}, \bibinfo {author} {\bibfnamefont
  {K.}~\bibnamefont {Watanabe}}, \ and\ \bibinfo {author} {\bibfnamefont
  {D.~K.}\ \bibnamefont {Bediako}},\ }\href
  {https://www.nature.com/articles/s41563-021-00973-w} {\bibfield  {journal}
  {\bibinfo  {journal} {Nat. Mater.}\ }\textbf {\bibinfo {volume} {20}},\
  \bibinfo {pages} {956} (\bibinfo {year} {2021})}\BibitemShut {NoStop}%
\bibitem [{\citenamefont {Saito}\ \emph {et~al.}(2021)\citenamefont {Saito},
  \citenamefont {Yang}, \citenamefont {Ge}, \citenamefont {Liu}, \citenamefont
  {Taniguchi}, \citenamefont {Watanabe}, \citenamefont {Li}, \citenamefont
  {Berg},\ and\ \citenamefont {Young}}]{saito2021isospin}%
  \BibitemOpen
  \bibfield  {author} {\bibinfo {author} {\bibfnamefont {Y.}~\bibnamefont
  {Saito}}, \bibinfo {author} {\bibfnamefont {F.}~\bibnamefont {Yang}},
  \bibinfo {author} {\bibfnamefont {J.}~\bibnamefont {Ge}}, \bibinfo {author}
  {\bibfnamefont {X.}~\bibnamefont {Liu}}, \bibinfo {author} {\bibfnamefont
  {T.}~\bibnamefont {Taniguchi}}, \bibinfo {author} {\bibfnamefont
  {K.}~\bibnamefont {Watanabe}}, \bibinfo {author} {\bibfnamefont
  {J.}~\bibnamefont {Li}}, \bibinfo {author} {\bibfnamefont {E.}~\bibnamefont
  {Berg}}, \ and\ \bibinfo {author} {\bibfnamefont {A.~F.}\ \bibnamefont
  {Young}},\ }\href {https://www.nature.com/articles/s41586-021-03409-2}
  {\bibfield  {journal} {\bibinfo  {journal} {Nature}\ }\textbf {\bibinfo
  {volume} {592}},\ \bibinfo {pages} {220} (\bibinfo {year}
  {2021})}\BibitemShut {NoStop}%
\bibitem [{\citenamefont {Wong}\ \emph {et~al.}(2020)\citenamefont {Wong},
  \citenamefont {Nuckolls}, \citenamefont {Oh}, \citenamefont {Lian},
  \citenamefont {Xie}, \citenamefont {Jeon}, \citenamefont {Watanabe},
  \citenamefont {Taniguchi}, \citenamefont {Bernevig},\ and\ \citenamefont
  {Yazdani}}]{wong2020cascade}%
  \BibitemOpen
  \bibfield  {author} {\bibinfo {author} {\bibfnamefont {D.}~\bibnamefont
  {Wong}}, \bibinfo {author} {\bibfnamefont {K.~P.}\ \bibnamefont {Nuckolls}},
  \bibinfo {author} {\bibfnamefont {M.}~\bibnamefont {Oh}}, \bibinfo {author}
  {\bibfnamefont {B.}~\bibnamefont {Lian}}, \bibinfo {author} {\bibfnamefont
  {Y.}~\bibnamefont {Xie}}, \bibinfo {author} {\bibfnamefont {S.}~\bibnamefont
  {Jeon}}, \bibinfo {author} {\bibfnamefont {K.}~\bibnamefont {Watanabe}},
  \bibinfo {author} {\bibfnamefont {T.}~\bibnamefont {Taniguchi}}, \bibinfo
  {author} {\bibfnamefont {B.~A.}\ \bibnamefont {Bernevig}}, \ and\ \bibinfo
  {author} {\bibfnamefont {A.}~\bibnamefont {Yazdani}},\ }\href
  {https://www.nature.com/articles/s41586-020-2339-0} {\bibfield  {journal}
  {\bibinfo  {journal} {Nature}\ }\textbf {\bibinfo {volume} {582}},\ \bibinfo
  {pages} {198} (\bibinfo {year} {2020})}\BibitemShut {NoStop}%
\bibitem [{\citenamefont {Oh}\ \emph {et~al.}(2021)\citenamefont {Oh},
  \citenamefont {Nuckolls}, \citenamefont {Wong}, \citenamefont {Lee},
  \citenamefont {Liu}, \citenamefont {Watanabe}, \citenamefont {Taniguchi},\
  and\ \citenamefont {Yazdani}}]{oh2021evidence}%
  \BibitemOpen
  \bibfield  {author} {\bibinfo {author} {\bibfnamefont {M.}~\bibnamefont
  {Oh}}, \bibinfo {author} {\bibfnamefont {K.~P.}\ \bibnamefont {Nuckolls}},
  \bibinfo {author} {\bibfnamefont {D.}~\bibnamefont {Wong}}, \bibinfo {author}
  {\bibfnamefont {R.~L.}\ \bibnamefont {Lee}}, \bibinfo {author} {\bibfnamefont
  {X.}~\bibnamefont {Liu}}, \bibinfo {author} {\bibfnamefont {K.}~\bibnamefont
  {Watanabe}}, \bibinfo {author} {\bibfnamefont {T.}~\bibnamefont {Taniguchi}},
  \ and\ \bibinfo {author} {\bibfnamefont {A.}~\bibnamefont {Yazdani}},\ }\href
  {https://www.nature.com/articles/s41586-021-04121-x} {\bibfield  {journal}
  {\bibinfo  {journal} {Nature}\ }\textbf {\bibinfo {volume} {600}},\ \bibinfo
  {pages} {240} (\bibinfo {year} {2021})}\BibitemShut {NoStop}%
\bibitem [{\citenamefont {Yoo}\ \emph {et~al.}(2019)\citenamefont {Yoo},
  \citenamefont {Engelke}, \citenamefont {Carr}, \citenamefont {Fang},
  \citenamefont {Zhang}, \citenamefont {Cazeaux}, \citenamefont {Sung},
  \citenamefont {Hovden}, \citenamefont {Tsen}, \citenamefont {Taniguchi} \emph
  {et~al.}}]{yoo2019atomic}%
  \BibitemOpen
  \bibfield  {author} {\bibinfo {author} {\bibfnamefont {H.}~\bibnamefont
  {Yoo}}, \bibinfo {author} {\bibfnamefont {R.}~\bibnamefont {Engelke}},
  \bibinfo {author} {\bibfnamefont {S.}~\bibnamefont {Carr}}, \bibinfo {author}
  {\bibfnamefont {S.}~\bibnamefont {Fang}}, \bibinfo {author} {\bibfnamefont
  {K.}~\bibnamefont {Zhang}}, \bibinfo {author} {\bibfnamefont
  {P.}~\bibnamefont {Cazeaux}}, \bibinfo {author} {\bibfnamefont {S.~H.}\
  \bibnamefont {Sung}}, \bibinfo {author} {\bibfnamefont {R.}~\bibnamefont
  {Hovden}}, \bibinfo {author} {\bibfnamefont {A.~W.}\ \bibnamefont {Tsen}},
  \bibinfo {author} {\bibfnamefont {T.}~\bibnamefont {Taniguchi}},  \emph
  {et~al.},\ }\href {https://www.nature.com/articles/s41563-019-0346-z}
  {\bibfield  {journal} {\bibinfo  {journal} {Nat. Mater.}\ }\textbf {\bibinfo
  {volume} {18}},\ \bibinfo {pages} {448} (\bibinfo {year} {2019})}\BibitemShut
  {NoStop}%
\bibitem [{\citenamefont {Gadelha}\ \emph {et~al.}(2021)\citenamefont
  {Gadelha}, \citenamefont {Ohlberg}, \citenamefont {Rabelo}, \citenamefont
  {Neto}, \citenamefont {Vasconcelos}, \citenamefont {Campos}, \citenamefont
  {Lemos}, \citenamefont {Ornelas}, \citenamefont {Miranda}, \citenamefont
  {Nadas} \emph {et~al.}}]{gadelha2021localization}%
  \BibitemOpen
  \bibfield  {author} {\bibinfo {author} {\bibfnamefont {A.~C.}\ \bibnamefont
  {Gadelha}}, \bibinfo {author} {\bibfnamefont {D.~A.}\ \bibnamefont
  {Ohlberg}}, \bibinfo {author} {\bibfnamefont {C.}~\bibnamefont {Rabelo}},
  \bibinfo {author} {\bibfnamefont {E.~G.}\ \bibnamefont {Neto}}, \bibinfo
  {author} {\bibfnamefont {T.~L.}\ \bibnamefont {Vasconcelos}}, \bibinfo
  {author} {\bibfnamefont {J.~L.}\ \bibnamefont {Campos}}, \bibinfo {author}
  {\bibfnamefont {J.~S.}\ \bibnamefont {Lemos}}, \bibinfo {author}
  {\bibfnamefont {V.}~\bibnamefont {Ornelas}}, \bibinfo {author} {\bibfnamefont
  {D.}~\bibnamefont {Miranda}}, \bibinfo {author} {\bibfnamefont
  {R.}~\bibnamefont {Nadas}},  \emph {et~al.},\ }\href
  {https://www.nature.com/articles/s41586-021-03252-5} {\bibfield  {journal}
  {\bibinfo  {journal} {Nature}\ }\textbf {\bibinfo {volume} {590}},\ \bibinfo
  {pages} {405} (\bibinfo {year} {2021})}\BibitemShut {NoStop}%
\bibitem [{\citenamefont {Jiang}\ \emph {et~al.}(2019)\citenamefont {Jiang},
  \citenamefont {Lai}, \citenamefont {Watanabe}, \citenamefont {Taniguchi},
  \citenamefont {Haule}, \citenamefont {Mao},\ and\ \citenamefont
  {Andrei}}]{jiang2019charge}%
  \BibitemOpen
  \bibfield  {author} {\bibinfo {author} {\bibfnamefont {Y.}~\bibnamefont
  {Jiang}}, \bibinfo {author} {\bibfnamefont {X.}~\bibnamefont {Lai}}, \bibinfo
  {author} {\bibfnamefont {K.}~\bibnamefont {Watanabe}}, \bibinfo {author}
  {\bibfnamefont {T.}~\bibnamefont {Taniguchi}}, \bibinfo {author}
  {\bibfnamefont {K.}~\bibnamefont {Haule}}, \bibinfo {author} {\bibfnamefont
  {J.}~\bibnamefont {Mao}}, \ and\ \bibinfo {author} {\bibfnamefont {E.~Y.}\
  \bibnamefont {Andrei}},\ }\href
  {https://www.nature.com/articles/s41586-019-1460-4} {\bibfield  {journal}
  {\bibinfo  {journal} {Nature}\ }\textbf {\bibinfo {volume} {573}},\ \bibinfo
  {pages} {91} (\bibinfo {year} {2019})}\BibitemShut {NoStop}%
\bibitem [{\citenamefont {Xie}\ \emph {et~al.}(2019)\citenamefont {Xie},
  \citenamefont {Lian}, \citenamefont {J{\"a}ck}, \citenamefont {Liu},
  \citenamefont {Chiu}, \citenamefont {Watanabe}, \citenamefont {Taniguchi},
  \citenamefont {Bernevig},\ and\ \citenamefont
  {Yazdani}}]{xie2019spectroscopic}%
  \BibitemOpen
  \bibfield  {author} {\bibinfo {author} {\bibfnamefont {Y.}~\bibnamefont
  {Xie}}, \bibinfo {author} {\bibfnamefont {B.}~\bibnamefont {Lian}}, \bibinfo
  {author} {\bibfnamefont {B.}~\bibnamefont {J{\"a}ck}}, \bibinfo {author}
  {\bibfnamefont {X.}~\bibnamefont {Liu}}, \bibinfo {author} {\bibfnamefont
  {C.-L.}\ \bibnamefont {Chiu}}, \bibinfo {author} {\bibfnamefont
  {K.}~\bibnamefont {Watanabe}}, \bibinfo {author} {\bibfnamefont
  {T.}~\bibnamefont {Taniguchi}}, \bibinfo {author} {\bibfnamefont {B.~A.}\
  \bibnamefont {Bernevig}}, \ and\ \bibinfo {author} {\bibfnamefont
  {A.}~\bibnamefont {Yazdani}},\ }\href
  {https://www.nature.com/articles/s41586-019-1422-x} {\bibfield  {journal}
  {\bibinfo  {journal} {Nature}\ }\textbf {\bibinfo {volume} {572}},\ \bibinfo
  {pages} {101} (\bibinfo {year} {2019})}\BibitemShut {NoStop}%
\bibitem [{\citenamefont {Liu}\ \emph {et~al.}(2021)\citenamefont {Liu},
  \citenamefont {Wang}, \citenamefont {Watanabe}, \citenamefont {Taniguchi},
  \citenamefont {Vafek},\ and\ \citenamefont {Li}}]{liu2021tuning}%
  \BibitemOpen
  \bibfield  {author} {\bibinfo {author} {\bibfnamefont {X.}~\bibnamefont
  {Liu}}, \bibinfo {author} {\bibfnamefont {Z.}~\bibnamefont {Wang}}, \bibinfo
  {author} {\bibfnamefont {K.}~\bibnamefont {Watanabe}}, \bibinfo {author}
  {\bibfnamefont {T.}~\bibnamefont {Taniguchi}}, \bibinfo {author}
  {\bibfnamefont {O.}~\bibnamefont {Vafek}}, \ and\ \bibinfo {author}
  {\bibfnamefont {J.}~\bibnamefont {Li}},\ }\href
  {https://www.science.org/doi/10.1126/science.abb8754} {\bibfield  {journal}
  {\bibinfo  {journal} {Science}\ }\textbf {\bibinfo {volume} {371}},\ \bibinfo
  {pages} {1261} (\bibinfo {year} {2021})}\BibitemShut {NoStop}%
\bibitem [{\citenamefont {{Park}}\ \emph {et~al.}(2022)\citenamefont {{Park}},
  \citenamefont {{Cao}}, \citenamefont {{Xia}}, \citenamefont {{Sun}},
  \citenamefont {{Watanabe}}, \citenamefont {{Taniguchi}},\ and\ \citenamefont
  {{Jarillo-Herrero}}}]{ParkNatMa2022}%
  \BibitemOpen
  \bibfield  {author} {\bibinfo {author} {\bibfnamefont {J.~M.}\ \bibnamefont
  {{Park}}}, \bibinfo {author} {\bibfnamefont {Y.}~\bibnamefont {{Cao}}},
  \bibinfo {author} {\bibfnamefont {L.-Q.}\ \bibnamefont {{Xia}}}, \bibinfo
  {author} {\bibfnamefont {S.}~\bibnamefont {{Sun}}}, \bibinfo {author}
  {\bibfnamefont {K.}~\bibnamefont {{Watanabe}}}, \bibinfo {author}
  {\bibfnamefont {T.}~\bibnamefont {{Taniguchi}}}, \ and\ \bibinfo {author}
  {\bibfnamefont {P.}~\bibnamefont {{Jarillo-Herrero}}},\ }\href {\doibase
  10.1038/s41563-022-01287-1} {\bibfield  {journal} {\bibinfo  {journal} {Nat.
  Mater.}\ }\textbf {\bibinfo {volume} {21}},\ \bibinfo {pages} {877} (\bibinfo
  {year} {2022})}\BibitemShut {NoStop}%
\bibitem [{\citenamefont {Burg}\ \emph {et~al.}(2022)\citenamefont {Burg},
  \citenamefont {Khalaf}, \citenamefont {Wang}, \citenamefont {Watanabe},
  \citenamefont {Taniguchi},\ and\ \citenamefont
  {Tutuc}}]{fourlayer2022natmaterial}%
  \BibitemOpen
  \bibfield  {author} {\bibinfo {author} {\bibfnamefont {G.}~\bibnamefont
  {Burg}}, \bibinfo {author} {\bibfnamefont {E.}~\bibnamefont {Khalaf}},
  \bibinfo {author} {\bibfnamefont {Y.}~\bibnamefont {Wang}}, \bibinfo {author}
  {\bibfnamefont {K.}~\bibnamefont {Watanabe}}, \bibinfo {author}
  {\bibfnamefont {T.}~\bibnamefont {Taniguchi}}, \ and\ \bibinfo {author}
  {\bibfnamefont {E.}~\bibnamefont {Tutuc}},\ }\href {\doibase
  10.1038/s41563-022-01286-2} {\bibfield  {journal} {\bibinfo  {journal} {Nat.
  Mater.}\ }\textbf {\bibinfo {volume} {21}},\ \bibinfo {pages} {884} (\bibinfo
  {year} {2022})}\BibitemShut {NoStop}%
\bibitem [{\citenamefont {Shin}\ \emph {et~al.}(2022)\citenamefont {Shin},
  \citenamefont {Chittari}, \citenamefont {Jang}, \citenamefont {Min},\ and\
  \citenamefont {Jung}}]{jj2022prb}%
  \BibitemOpen
  \bibfield  {author} {\bibinfo {author} {\bibfnamefont {J.}~\bibnamefont
  {Shin}}, \bibinfo {author} {\bibfnamefont {B.~L.}\ \bibnamefont {Chittari}},
  \bibinfo {author} {\bibfnamefont {Y.}~\bibnamefont {Jang}}, \bibinfo {author}
  {\bibfnamefont {H.}~\bibnamefont {Min}}, \ and\ \bibinfo {author}
  {\bibfnamefont {J.}~\bibnamefont {Jung}},\ }\href {\doibase
  10.1103/PhysRevB.105.245124} {\bibfield  {journal} {\bibinfo  {journal}
  {Phys. Rev. B}\ }\textbf {\bibinfo {volume} {105}},\ \bibinfo {pages}
  {245124} (\bibinfo {year} {2022})}\BibitemShut {NoStop}%
\bibitem [{\citenamefont {Liu}\ \emph {et~al.}(2022)\citenamefont {Liu},
  \citenamefont {Shi}, \citenamefont {Yang},\ and\ \citenamefont
  {Zhang}}]{LIU202228}%
  \BibitemOpen
  \bibfield  {author} {\bibinfo {author} {\bibfnamefont {Z.}~\bibnamefont
  {Liu}}, \bibinfo {author} {\bibfnamefont {W.}~\bibnamefont {Shi}}, \bibinfo
  {author} {\bibfnamefont {T.}~\bibnamefont {Yang}}, \ and\ \bibinfo {author}
  {\bibfnamefont {Z.}~\bibnamefont {Zhang}},\ }\href {\doibase
  https://doi.org/10.1016/j.jmst.2021.09.040} {\bibfield  {journal} {\bibinfo
  {journal} {J. Mater. Sci. Technol.}\ }\textbf {\bibinfo {volume} {111}},\
  \bibinfo {pages} {28} (\bibinfo {year} {2022})}\BibitemShut {NoStop}%
\bibitem [{\citenamefont {Liang}\ \emph {et~al.}(2022)\citenamefont {Liang},
  \citenamefont {Xiao}, \citenamefont {Ma},\ and\ \citenamefont
  {Gao}}]{Liangmiao2022prb}%
  \BibitemOpen
  \bibfield  {author} {\bibinfo {author} {\bibfnamefont {M.}~\bibnamefont
  {Liang}}, \bibinfo {author} {\bibfnamefont {M.-M.}\ \bibnamefont {Xiao}},
  \bibinfo {author} {\bibfnamefont {Z.}~\bibnamefont {Ma}}, \ and\ \bibinfo
  {author} {\bibfnamefont {J.-H.}\ \bibnamefont {Gao}},\ }\href {\doibase
  10.1103/PhysRevB.105.195422} {\bibfield  {journal} {\bibinfo  {journal}
  {Phys. Rev. B}\ }\textbf {\bibinfo {volume} {105}},\ \bibinfo {pages}
  {195422} (\bibinfo {year} {2022})}\BibitemShut {NoStop}%
\bibitem [{\citenamefont {{Ma}}\ \emph {et~al.}(2023)\citenamefont {{Ma}},
  \citenamefont {{Li}}, \citenamefont {{Lu}}, \citenamefont {{Xu}},
  \citenamefont {{Gao}},\ and\ \citenamefont {{Xie}}}]{mazhen2023}%
  \BibitemOpen
  \bibfield  {author} {\bibinfo {author} {\bibfnamefont {Z.}~\bibnamefont
  {{Ma}}}, \bibinfo {author} {\bibfnamefont {S.}~\bibnamefont {{Li}}}, \bibinfo
  {author} {\bibfnamefont {M.}~\bibnamefont {{Lu}}}, \bibinfo {author}
  {\bibfnamefont {D.-H.}\ \bibnamefont {{Xu}}}, \bibinfo {author}
  {\bibfnamefont {J.-H.}\ \bibnamefont {{Gao}}}, \ and\ \bibinfo {author}
  {\bibfnamefont {X.}~\bibnamefont {{Xie}}},\ }\href {\doibase
  10.1007/s11433-022-1993-7} {\bibfield  {journal} {\bibinfo  {journal} {Sci.
  China Phys. Mech. Astron.}\ }\textbf {\bibinfo {volume} {66}},\ \bibinfo
  {eid} {227211} (\bibinfo {year} {2023})}\BibitemShut {NoStop}%
\bibitem [{\citenamefont {Xie}\ \emph {et~al.}(2022)\citenamefont {Xie},
  \citenamefont {Peng}, \citenamefont {Zhang},\ and\ \citenamefont
  {Liu}}]{xie2022alternating}%
  \BibitemOpen
  \bibfield  {author} {\bibinfo {author} {\bibfnamefont {B.}~\bibnamefont
  {Xie}}, \bibinfo {author} {\bibfnamefont {R.}~\bibnamefont {Peng}}, \bibinfo
  {author} {\bibfnamefont {S.}~\bibnamefont {Zhang}}, \ and\ \bibinfo {author}
  {\bibfnamefont {J.}~\bibnamefont {Liu}},\ }\href {\doibase
  10.1038/s41524-022-00789-5} {\bibfield  {journal} {\bibinfo  {journal} {npj
  Comput. Mater.}\ }\textbf {\bibinfo {volume} {8}},\ \bibinfo {pages} {1}
  (\bibinfo {year} {2022})}\BibitemShut {NoStop}%
\bibitem [{\citenamefont {Mora}\ \emph {et~al.}(2019)\citenamefont {Mora},
  \citenamefont {Regnault},\ and\ \citenamefont {Bernevig}}]{mora2019prl}%
  \BibitemOpen
  \bibfield  {author} {\bibinfo {author} {\bibfnamefont {C.}~\bibnamefont
  {Mora}}, \bibinfo {author} {\bibfnamefont {N.}~\bibnamefont {Regnault}}, \
  and\ \bibinfo {author} {\bibfnamefont {B.~A.}\ \bibnamefont {Bernevig}},\
  }\href {\doibase 10.1103/PhysRevLett.123.026402} {\bibfield  {journal}
  {\bibinfo  {journal} {Phys. Rev. Lett.}\ }\textbf {\bibinfo {volume} {123}},\
  \bibinfo {pages} {026402} (\bibinfo {year} {2019})}\BibitemShut {NoStop}%
\bibitem [{\citenamefont {Zhang}\ \emph {et~al.}(2021)\citenamefont {Zhang},
  \citenamefont {Tsai}, \citenamefont {Zhu}, \citenamefont {Ren}, \citenamefont
  {Luo}, \citenamefont {Carr}, \citenamefont {Luskin}, \citenamefont
  {Kaxiras},\ and\ \citenamefont {Wang}}]{wangke2021prl}%
  \BibitemOpen
  \bibfield  {author} {\bibinfo {author} {\bibfnamefont {X.}~\bibnamefont
  {Zhang}}, \bibinfo {author} {\bibfnamefont {K.-T.}\ \bibnamefont {Tsai}},
  \bibinfo {author} {\bibfnamefont {Z.}~\bibnamefont {Zhu}}, \bibinfo {author}
  {\bibfnamefont {W.}~\bibnamefont {Ren}}, \bibinfo {author} {\bibfnamefont
  {Y.}~\bibnamefont {Luo}}, \bibinfo {author} {\bibfnamefont {S.}~\bibnamefont
  {Carr}}, \bibinfo {author} {\bibfnamefont {M.}~\bibnamefont {Luskin}},
  \bibinfo {author} {\bibfnamefont {E.}~\bibnamefont {Kaxiras}}, \ and\
  \bibinfo {author} {\bibfnamefont {K.}~\bibnamefont {Wang}},\ }\href {\doibase
  10.1103/PhysRevLett.127.166802} {\bibfield  {journal} {\bibinfo  {journal}
  {Phys. Rev. Lett.}\ }\textbf {\bibinfo {volume} {127}},\ \bibinfo {pages}
  {166802} (\bibinfo {year} {2021})}\BibitemShut {NoStop}%
\bibitem [{\citenamefont {Zhu}\ \emph {et~al.}(2020)\citenamefont {Zhu},
  \citenamefont {Carr}, \citenamefont {Massatt}, \citenamefont {Luskin},\ and\
  \citenamefont {Kaxiras}}]{zhuziyan2020prl}%
  \BibitemOpen
  \bibfield  {author} {\bibinfo {author} {\bibfnamefont {Z.}~\bibnamefont
  {Zhu}}, \bibinfo {author} {\bibfnamefont {S.}~\bibnamefont {Carr}}, \bibinfo
  {author} {\bibfnamefont {D.}~\bibnamefont {Massatt}}, \bibinfo {author}
  {\bibfnamefont {M.}~\bibnamefont {Luskin}}, \ and\ \bibinfo {author}
  {\bibfnamefont {E.}~\bibnamefont {Kaxiras}},\ }\href {\doibase
  10.1103/PhysRevLett.125.116404} {\bibfield  {journal} {\bibinfo  {journal}
  {Phys. Rev. Lett.}\ }\textbf {\bibinfo {volume} {125}},\ \bibinfo {pages}
  {116404} (\bibinfo {year} {2020})}\BibitemShut {NoStop}%
\bibitem [{\citenamefont {Bistritzer}\ and\ \citenamefont
  {MacDonald}(2011)}]{pnas2011}%
  \BibitemOpen
  \bibfield  {author} {\bibinfo {author} {\bibfnamefont {R.}~\bibnamefont
  {Bistritzer}}\ and\ \bibinfo {author} {\bibfnamefont {A.~H.}\ \bibnamefont
  {MacDonald}},\ }\href {\doibase 10.1073/pnas.1108174108} {\bibfield
  {journal} {\bibinfo  {journal} {Proc. Natl. Acad. Sci.}\ }\textbf {\bibinfo
  {volume} {108}},\ \bibinfo {pages} {12233} (\bibinfo {year}
  {2011})}\BibitemShut {NoStop}%
\bibitem [{\citenamefont {Moon}\ and\ \citenamefont
  {Koshino}(2013)}]{moon2013prb}%
  \BibitemOpen
  \bibfield  {author} {\bibinfo {author} {\bibfnamefont {P.}~\bibnamefont
  {Moon}}\ and\ \bibinfo {author} {\bibfnamefont {M.}~\bibnamefont {Koshino}},\
  }\href {\doibase 10.1103/PhysRevB.87.205404} {\bibfield  {journal} {\bibinfo
  {journal} {Phys. Rev. B}\ }\textbf {\bibinfo {volume} {87}},\ \bibinfo
  {pages} {205404} (\bibinfo {year} {2013})}\BibitemShut {NoStop}%
\bibitem [{\citenamefont {Koshino}(2015)}]{Koshino_2015}%
  \BibitemOpen
  \bibfield  {author} {\bibinfo {author} {\bibfnamefont {M.}~\bibnamefont
  {Koshino}},\ }\href {\doibase 10.1088/1367-2630/17/1/015014} {\bibfield
  {journal} {\bibinfo  {journal} {New J. Phys.}\ }\textbf {\bibinfo {volume}
  {17}},\ \bibinfo {pages} {015014} (\bibinfo {year} {2015})}\BibitemShut
  {NoStop}%
\bibitem [{\citenamefont {Koshino}\ and\ \citenamefont
  {Moon}(2015)}]{moon2015}%
  \BibitemOpen
  \bibfield  {author} {\bibinfo {author} {\bibfnamefont {M.}~\bibnamefont
  {Koshino}}\ and\ \bibinfo {author} {\bibfnamefont {P.}~\bibnamefont {Moon}},\
  }\href {\doibase 10.7566/JPSJ.84.121001} {\bibfield  {journal} {\bibinfo
  {journal} {J. Phys. Soc. Jpn.}\ }\textbf {\bibinfo {volume} {84}},\ \bibinfo
  {pages} {121001} (\bibinfo {year} {2015})}\BibitemShut {NoStop}%
\bibitem [{\citenamefont {Ma}\ \emph {et~al.}(2023{\natexlab{a}})\citenamefont
  {Ma}, \citenamefont {Li}, \citenamefont {Xiao}, \citenamefont {Zheng},
  \citenamefont {Lu}, \citenamefont {Liu}, \citenamefont {Gao},\ and\
  \citenamefont {Xie}}]{Ma2020front}%
  \BibitemOpen
  \bibfield  {author} {\bibinfo {author} {\bibfnamefont {Z.}~\bibnamefont
  {Ma}}, \bibinfo {author} {\bibfnamefont {S.}~\bibnamefont {Li}}, \bibinfo
  {author} {\bibfnamefont {M.-M.}\ \bibnamefont {Xiao}}, \bibinfo {author}
  {\bibfnamefont {Y.-W.}\ \bibnamefont {Zheng}}, \bibinfo {author}
  {\bibfnamefont {M.}~\bibnamefont {Lu}}, \bibinfo {author} {\bibfnamefont
  {H.}~\bibnamefont {Liu}}, \bibinfo {author} {\bibfnamefont {J.-H.}\
  \bibnamefont {Gao}}, \ and\ \bibinfo {author} {\bibfnamefont
  {X.}~\bibnamefont {Xie}},\ }\href {\doibase 10.48550/arXiv.2001.07995}
  {\bibfield  {journal} {\bibinfo  {journal} {Front. Phys.}\ }\textbf {\bibinfo
  {volume} {18}},\ \bibinfo {pages} {1} (\bibinfo {year}
  {2023}{\natexlab{a}})}\BibitemShut {NoStop}%
\bibitem [{\citenamefont {Koshino}\ and\ \citenamefont
  {McCann}(2009)}]{koshino2009prb}%
  \BibitemOpen
  \bibfield  {author} {\bibinfo {author} {\bibfnamefont {M.}~\bibnamefont
  {Koshino}}\ and\ \bibinfo {author} {\bibfnamefont {E.}~\bibnamefont
  {McCann}},\ }\href {\doibase 10.1103/PhysRevB.79.125443} {\bibfield
  {journal} {\bibinfo  {journal} {Phys. Rev. B}\ }\textbf {\bibinfo {volume}
  {79}},\ \bibinfo {pages} {125443} (\bibinfo {year} {2009})}\BibitemShut
  {NoStop}%
\bibitem [{\citenamefont {Stepanov}\ \emph {et~al.}(2019)\citenamefont
  {Stepanov}, \citenamefont {Barlas}, \citenamefont {Che}, \citenamefont
  {Myhro}, \citenamefont {Voigt}, \citenamefont {Pi}, \citenamefont {Watanabe},
  \citenamefont {Taniguchi}, \citenamefont {Smirnov}, \citenamefont {Zhang},
  \citenamefont {Lake}, \citenamefont {MacDonald},\ and\ \citenamefont
  {Lau}}]{zhangfan2019pnas}%
  \BibitemOpen
  \bibfield  {author} {\bibinfo {author} {\bibfnamefont {P.}~\bibnamefont
  {Stepanov}}, \bibinfo {author} {\bibfnamefont {Y.}~\bibnamefont {Barlas}},
  \bibinfo {author} {\bibfnamefont {S.}~\bibnamefont {Che}}, \bibinfo {author}
  {\bibfnamefont {K.}~\bibnamefont {Myhro}}, \bibinfo {author} {\bibfnamefont
  {G.}~\bibnamefont {Voigt}}, \bibinfo {author} {\bibfnamefont
  {Z.}~\bibnamefont {Pi}}, \bibinfo {author} {\bibfnamefont {K.}~\bibnamefont
  {Watanabe}}, \bibinfo {author} {\bibfnamefont {T.}~\bibnamefont {Taniguchi}},
  \bibinfo {author} {\bibfnamefont {D.}~\bibnamefont {Smirnov}}, \bibinfo
  {author} {\bibfnamefont {F.}~\bibnamefont {Zhang}}, \bibinfo {author}
  {\bibfnamefont {R.~K.}\ \bibnamefont {Lake}}, \bibinfo {author}
  {\bibfnamefont {A.~H.}\ \bibnamefont {MacDonald}}, \ and\ \bibinfo {author}
  {\bibfnamefont {C.~N.}\ \bibnamefont {Lau}},\ }\href
  {https://www.pnas.org/doi/10.1073/pnas.1820835116} {\bibfield  {journal}
  {\bibinfo  {journal} {Proc. Natl. Acad. Sci.}\ }\textbf {\bibinfo {volume}
  {116}},\ \bibinfo {pages} {10286} (\bibinfo {year} {2019})}\BibitemShut
  {NoStop}%
\bibitem [{\citenamefont {Ma}\ \emph {et~al.}(2021)\citenamefont {Ma},
  \citenamefont {Li}, \citenamefont {Zheng}, \citenamefont {Xiao},
  \citenamefont {Jiang}, \citenamefont {Gao},\ and\ \citenamefont
  {Xie}}]{MA202118}%
  \BibitemOpen
  \bibfield  {author} {\bibinfo {author} {\bibfnamefont {Z.}~\bibnamefont
  {Ma}}, \bibinfo {author} {\bibfnamefont {S.}~\bibnamefont {Li}}, \bibinfo
  {author} {\bibfnamefont {Y.-W.}\ \bibnamefont {Zheng}}, \bibinfo {author}
  {\bibfnamefont {M.-M.}\ \bibnamefont {Xiao}}, \bibinfo {author}
  {\bibfnamefont {H.}~\bibnamefont {Jiang}}, \bibinfo {author} {\bibfnamefont
  {J.-H.}\ \bibnamefont {Gao}}, \ and\ \bibinfo {author} {\bibfnamefont
  {X.}~\bibnamefont {Xie}},\ }\href {\doibase
  https://doi.org/10.1016/j.scib.2020.10.004} {\bibfield  {journal} {\bibinfo
  {journal} {Sci. Bull.}\ }\textbf {\bibinfo {volume} {66}},\ \bibinfo {pages}
  {18} (\bibinfo {year} {2021})}\BibitemShut {NoStop}%
\bibitem [{\citenamefont {Park}\ \emph {et~al.}(2020)\citenamefont {Park},
  \citenamefont {Chittari},\ and\ \citenamefont {Jung}}]{jj2020prb}%
  \BibitemOpen
  \bibfield  {author} {\bibinfo {author} {\bibfnamefont {Y.}~\bibnamefont
  {Park}}, \bibinfo {author} {\bibfnamefont {B.~L.}\ \bibnamefont {Chittari}},
  \ and\ \bibinfo {author} {\bibfnamefont {J.}~\bibnamefont {Jung}},\ }\href
  {\doibase 10.1103/PhysRevB.102.035411} {\bibfield  {journal} {\bibinfo
  {journal} {Phys. Rev. B}\ }\textbf {\bibinfo {volume} {102}},\ \bibinfo
  {pages} {035411} (\bibinfo {year} {2020})}\BibitemShut {NoStop}%
\bibitem [{\citenamefont {Rademaker}\ \emph {et~al.}(2020)\citenamefont
  {Rademaker}, \citenamefont {Protopopov},\ and\ \citenamefont
  {Abanin}}]{rademaker2020prr}%
  \BibitemOpen
  \bibfield  {author} {\bibinfo {author} {\bibfnamefont {L.}~\bibnamefont
  {Rademaker}}, \bibinfo {author} {\bibfnamefont {I.~V.}\ \bibnamefont
  {Protopopov}}, \ and\ \bibinfo {author} {\bibfnamefont {D.~A.}\ \bibnamefont
  {Abanin}},\ }\href {\doibase 10.1103/PhysRevResearch.2.033150} {\bibfield
  {journal} {\bibinfo  {journal} {Phys. Rev. Res.}\ }\textbf {\bibinfo {volume}
  {2}},\ \bibinfo {pages} {033150} (\bibinfo {year} {2020})}\BibitemShut
  {NoStop}%
\bibitem [{\citenamefont {Chen}\ \emph {et~al.}(2021)\citenamefont {Chen},
  \citenamefont {He}, \citenamefont {Zhang}, \citenamefont {Hsieh},
  \citenamefont {Fei}, \citenamefont {Watanabe}, \citenamefont {Taniguchi},
  \citenamefont {Cobden}, \citenamefont {Xu}, \citenamefont {Dean} \emph
  {et~al.}}]{chen2021TMBG}%
  \BibitemOpen
  \bibfield  {author} {\bibinfo {author} {\bibfnamefont {S.}~\bibnamefont
  {Chen}}, \bibinfo {author} {\bibfnamefont {M.}~\bibnamefont {He}}, \bibinfo
  {author} {\bibfnamefont {Y.-H.}\ \bibnamefont {Zhang}}, \bibinfo {author}
  {\bibfnamefont {V.}~\bibnamefont {Hsieh}}, \bibinfo {author} {\bibfnamefont
  {Z.}~\bibnamefont {Fei}}, \bibinfo {author} {\bibfnamefont {K.}~\bibnamefont
  {Watanabe}}, \bibinfo {author} {\bibfnamefont {T.}~\bibnamefont {Taniguchi}},
  \bibinfo {author} {\bibfnamefont {D.~H.}\ \bibnamefont {Cobden}}, \bibinfo
  {author} {\bibfnamefont {X.}~\bibnamefont {Xu}}, \bibinfo {author}
  {\bibfnamefont {C.~R.}\ \bibnamefont {Dean}},  \emph {et~al.},\ }\href
  {\doibase 10.1038/s41567-020-01062-6} {\bibfield  {journal} {\bibinfo
  {journal} {Nat. Phys.}\ }\textbf {\bibinfo {volume} {17}},\ \bibinfo {pages}
  {374} (\bibinfo {year} {2021})}\BibitemShut {NoStop}%
\bibitem [{\citenamefont {Xu}\ \emph {et~al.}(2021)\citenamefont {Xu},
  \citenamefont {Al~Ezzi}, \citenamefont {Balakrishnan}, \citenamefont
  {Garcia-Ruiz}, \citenamefont {Tsim}, \citenamefont {Mullan}, \citenamefont
  {Barrier}, \citenamefont {Xin}, \citenamefont {Piot}, \citenamefont
  {Taniguchi} \emph {et~al.}}]{xu2021tmbg}%
  \BibitemOpen
  \bibfield  {author} {\bibinfo {author} {\bibfnamefont {S.}~\bibnamefont
  {Xu}}, \bibinfo {author} {\bibfnamefont {M.~M.}\ \bibnamefont {Al~Ezzi}},
  \bibinfo {author} {\bibfnamefont {N.}~\bibnamefont {Balakrishnan}}, \bibinfo
  {author} {\bibfnamefont {A.}~\bibnamefont {Garcia-Ruiz}}, \bibinfo {author}
  {\bibfnamefont {B.}~\bibnamefont {Tsim}}, \bibinfo {author} {\bibfnamefont
  {C.}~\bibnamefont {Mullan}}, \bibinfo {author} {\bibfnamefont
  {J.}~\bibnamefont {Barrier}}, \bibinfo {author} {\bibfnamefont
  {N.}~\bibnamefont {Xin}}, \bibinfo {author} {\bibfnamefont {B.~A.}\
  \bibnamefont {Piot}}, \bibinfo {author} {\bibfnamefont {T.}~\bibnamefont
  {Taniguchi}},  \emph {et~al.},\ }\href {\doibase 10.1038/s41567-021-01172-9}
  {\bibfield  {journal} {\bibinfo  {journal} {Nat. Phys.}\ }\textbf {\bibinfo
  {volume} {17}},\ \bibinfo {pages} {619} (\bibinfo {year} {2021})}\BibitemShut
  {NoStop}%
\bibitem [{\citenamefont {Polshyn}\ \emph {et~al.}(2020)\citenamefont
  {Polshyn}, \citenamefont {Zhu}, \citenamefont {Kumar}, \citenamefont {Zhang},
  \citenamefont {Yang}, \citenamefont {Tschirhart}, \citenamefont {Serlin},
  \citenamefont {Watanabe}, \citenamefont {Taniguchi}, \citenamefont
  {MacDonald} \emph {et~al.}}]{polshyn2020tmbg}%
  \BibitemOpen
  \bibfield  {author} {\bibinfo {author} {\bibfnamefont {H.}~\bibnamefont
  {Polshyn}}, \bibinfo {author} {\bibfnamefont {J.}~\bibnamefont {Zhu}},
  \bibinfo {author} {\bibfnamefont {M.~A.}\ \bibnamefont {Kumar}}, \bibinfo
  {author} {\bibfnamefont {Y.}~\bibnamefont {Zhang}}, \bibinfo {author}
  {\bibfnamefont {F.}~\bibnamefont {Yang}}, \bibinfo {author} {\bibfnamefont
  {C.~L.}\ \bibnamefont {Tschirhart}}, \bibinfo {author} {\bibfnamefont
  {M.}~\bibnamefont {Serlin}}, \bibinfo {author} {\bibfnamefont
  {K.}~\bibnamefont {Watanabe}}, \bibinfo {author} {\bibfnamefont
  {T.}~\bibnamefont {Taniguchi}}, \bibinfo {author} {\bibfnamefont {A.~H.}\
  \bibnamefont {MacDonald}},  \emph {et~al.},\ }\href {\doibase
  10.1038/s41586-020-2963-8} {\bibfield  {journal} {\bibinfo  {journal}
  {Nature}\ }\textbf {\bibinfo {volume} {588}},\ \bibinfo {pages} {66}
  (\bibinfo {year} {2020})}\BibitemShut {NoStop}%
\bibitem [{\citenamefont {He}\ \emph {et~al.}(2021)\citenamefont {He},
  \citenamefont {Zhang}, \citenamefont {Li}, \citenamefont {Fei}, \citenamefont
  {Watanabe}, \citenamefont {Taniguchi}, \citenamefont {Xu},\ and\
  \citenamefont {Yankowitz}}]{he2021tmbg}%
  \BibitemOpen
  \bibfield  {author} {\bibinfo {author} {\bibfnamefont {M.}~\bibnamefont
  {He}}, \bibinfo {author} {\bibfnamefont {Y.-H.}\ \bibnamefont {Zhang}},
  \bibinfo {author} {\bibfnamefont {Y.}~\bibnamefont {Li}}, \bibinfo {author}
  {\bibfnamefont {Z.}~\bibnamefont {Fei}}, \bibinfo {author} {\bibfnamefont
  {K.}~\bibnamefont {Watanabe}}, \bibinfo {author} {\bibfnamefont
  {T.}~\bibnamefont {Taniguchi}}, \bibinfo {author} {\bibfnamefont
  {X.}~\bibnamefont {Xu}}, \ and\ \bibinfo {author} {\bibfnamefont
  {M.}~\bibnamefont {Yankowitz}},\ }\href {\doibase 10.1038/s41467-021-25044-1}
  {\bibfield  {journal} {\bibinfo  {journal} {Nat. Commun.}\ }\textbf {\bibinfo
  {volume} {12}},\ \bibinfo {pages} {1} (\bibinfo {year} {2021})}\BibitemShut
  {NoStop}%
\bibitem [{\citenamefont {Tong}\ \emph {et~al.}(2022)\citenamefont {Tong},
  \citenamefont {Tong}, \citenamefont {Yang}, \citenamefont {Zhou},
  \citenamefont {Wu}, \citenamefont {Tian}, \citenamefont {Zhang},
  \citenamefont {Zhang}, \citenamefont {Qin},\ and\ \citenamefont
  {Yin}}]{yin2022tmbg}%
  \BibitemOpen
  \bibfield  {author} {\bibinfo {author} {\bibfnamefont {L.-H.}\ \bibnamefont
  {Tong}}, \bibinfo {author} {\bibfnamefont {Q.}~\bibnamefont {Tong}}, \bibinfo
  {author} {\bibfnamefont {L.-Z.}\ \bibnamefont {Yang}}, \bibinfo {author}
  {\bibfnamefont {Y.-Y.}\ \bibnamefont {Zhou}}, \bibinfo {author}
  {\bibfnamefont {Q.}~\bibnamefont {Wu}}, \bibinfo {author} {\bibfnamefont
  {Y.}~\bibnamefont {Tian}}, \bibinfo {author} {\bibfnamefont {L.}~\bibnamefont
  {Zhang}}, \bibinfo {author} {\bibfnamefont {L.}~\bibnamefont {Zhang}},
  \bibinfo {author} {\bibfnamefont {Z.}~\bibnamefont {Qin}}, \ and\ \bibinfo
  {author} {\bibfnamefont {L.-J.}\ \bibnamefont {Yin}},\ }\href {\doibase
  10.1103/PhysRevLett.128.126401} {\bibfield  {journal} {\bibinfo  {journal}
  {Phys. Rev. Lett.}\ }\textbf {\bibinfo {volume} {128}},\ \bibinfo {pages}
  {126401} (\bibinfo {year} {2022})}\BibitemShut {NoStop}%
\bibitem [{\citenamefont {Ma}\ \emph {et~al.}(2023{\natexlab{b}})\citenamefont
  {Ma}, \citenamefont {Li}, \citenamefont {Lu}, \citenamefont {Xu},
  \citenamefont {Gao},\ and\ \citenamefont {Xie}}]{ma2021doubled}%
  \BibitemOpen
  \bibfield  {author} {\bibinfo {author} {\bibfnamefont {Z.}~\bibnamefont
  {Ma}}, \bibinfo {author} {\bibfnamefont {S.}~\bibnamefont {Li}}, \bibinfo
  {author} {\bibfnamefont {M.}~\bibnamefont {Lu}}, \bibinfo {author}
  {\bibfnamefont {D.-H.}\ \bibnamefont {Xu}}, \bibinfo {author} {\bibfnamefont
  {J.-H.}\ \bibnamefont {Gao}}, \ and\ \bibinfo {author} {\bibfnamefont
  {X.}~\bibnamefont {Xie}},\ }\href
  {https://link.springer.com/article/10.1007/s11433-022-1993-7} {\bibfield
  {journal} {\bibinfo  {journal} {Sci. China Phys. Mech. Astron.}\ }\textbf
  {\bibinfo {volume} {66}} (\bibinfo {year} {2023}{\natexlab{b}})}\BibitemShut
  {NoStop}%
\bibitem [{\citenamefont {Liu}\ \emph {et~al.}(2019)\citenamefont {Liu},
  \citenamefont {Ma}, \citenamefont {Gao},\ and\ \citenamefont
  {Dai}}]{liujianpengprx2019}%
  \BibitemOpen
  \bibfield  {author} {\bibinfo {author} {\bibfnamefont {J.}~\bibnamefont
  {Liu}}, \bibinfo {author} {\bibfnamefont {Z.}~\bibnamefont {Ma}}, \bibinfo
  {author} {\bibfnamefont {J.}~\bibnamefont {Gao}}, \ and\ \bibinfo {author}
  {\bibfnamefont {X.}~\bibnamefont {Dai}},\ }\href {\doibase
  10.1103/PhysRevX.9.031021} {\bibfield  {journal} {\bibinfo  {journal} {Phys.
  Rev. X}\ }\textbf {\bibinfo {volume} {9}},\ \bibinfo {pages} {031021}
  (\bibinfo {year} {2019})}\BibitemShut {NoStop}%
\bibitem [{\citenamefont {Zhang}\ \emph {et~al.}(2020)\citenamefont {Zhang},
  \citenamefont {Xie}, \citenamefont {Wu}, \citenamefont {Liu},\ and\
  \citenamefont {Yazyev}}]{zhang2020chiral}%
  \BibitemOpen
  \bibfield  {author} {\bibinfo {author} {\bibfnamefont {S.}~\bibnamefont
  {Zhang}}, \bibinfo {author} {\bibfnamefont {B.}~\bibnamefont {Xie}}, \bibinfo
  {author} {\bibfnamefont {Q.}~\bibnamefont {Wu}}, \bibinfo {author}
  {\bibfnamefont {J.}~\bibnamefont {Liu}}, \ and\ \bibinfo {author}
  {\bibfnamefont {O.~V.}\ \bibnamefont {Yazyev}},\ }\href {\doibase
  10.48550/arXiv.2012.11964} {\bibfield  {journal} {\bibinfo  {journal}
  {arXiv}\ } (\bibinfo {year} {2020}),\ 10.48550/arXiv.2012.11964}\BibitemShut
  {NoStop}%
\bibitem [{\citenamefont {Cao}\ \emph {et~al.}(2021{\natexlab{b}})\citenamefont
  {Cao}, \citenamefont {Wang}, \citenamefont {Qian}, \citenamefont {Liu},\ and\
  \citenamefont {Yao}}]{liuchengcheng2021prb}%
  \BibitemOpen
  \bibfield  {author} {\bibinfo {author} {\bibfnamefont {J.}~\bibnamefont
  {Cao}}, \bibinfo {author} {\bibfnamefont {M.}~\bibnamefont {Wang}}, \bibinfo
  {author} {\bibfnamefont {S.-F.}\ \bibnamefont {Qian}}, \bibinfo {author}
  {\bibfnamefont {C.-C.}\ \bibnamefont {Liu}}, \ and\ \bibinfo {author}
  {\bibfnamefont {Y.}~\bibnamefont {Yao}},\ }\href {\doibase
  10.1103/PhysRevB.104.L081403} {\bibfield  {journal} {\bibinfo  {journal}
  {Phys. Rev. B}\ }\textbf {\bibinfo {volume} {104}},\ \bibinfo {pages}
  {L081403} (\bibinfo {year} {2021}{\natexlab{b}})}\BibitemShut {NoStop}%
\bibitem [{\citenamefont {Cao}\ \emph {et~al.}(2020{\natexlab{b}})\citenamefont
  {Cao}, \citenamefont {Rodan-Legrain}, \citenamefont {Rubies-Bigorda},
  \citenamefont {Park}, \citenamefont {Watanabe}, \citenamefont {Taniguchi},\
  and\ \citenamefont {Jarillo-Herrero}}]{cao2020tunable}%
  \BibitemOpen
  \bibfield  {author} {\bibinfo {author} {\bibfnamefont {Y.}~\bibnamefont
  {Cao}}, \bibinfo {author} {\bibfnamefont {D.}~\bibnamefont {Rodan-Legrain}},
  \bibinfo {author} {\bibfnamefont {O.}~\bibnamefont {Rubies-Bigorda}},
  \bibinfo {author} {\bibfnamefont {J.~M.}\ \bibnamefont {Park}}, \bibinfo
  {author} {\bibfnamefont {K.}~\bibnamefont {Watanabe}}, \bibinfo {author}
  {\bibfnamefont {T.}~\bibnamefont {Taniguchi}}, \ and\ \bibinfo {author}
  {\bibfnamefont {P.}~\bibnamefont {Jarillo-Herrero}},\ }\href {\doibase
  10.1038/s41586-020-2260-6} {\bibfield  {journal} {\bibinfo  {journal}
  {Nature}\ }\textbf {\bibinfo {volume} {583}},\ \bibinfo {pages} {215}
  (\bibinfo {year} {2020}{\natexlab{b}})}\BibitemShut {NoStop}%
\bibitem [{\citenamefont {Liu}\ \emph {et~al.}(2020)\citenamefont {Liu},
  \citenamefont {Hao}, \citenamefont {Khalaf}, \citenamefont {Lee},
  \citenamefont {Ronen}, \citenamefont {Yoo}, \citenamefont {Najafabadi},
  \citenamefont {Watanabe}, \citenamefont {Taniguchi}, \citenamefont
  {Vishwanath},\ and\ \citenamefont {Kim}}]{Tunable583}%
  \BibitemOpen
  \bibfield  {author} {\bibinfo {author} {\bibfnamefont {X.}~\bibnamefont
  {Liu}}, \bibinfo {author} {\bibfnamefont {Z.}~\bibnamefont {Hao}}, \bibinfo
  {author} {\bibfnamefont {E.}~\bibnamefont {Khalaf}}, \bibinfo {author}
  {\bibfnamefont {J.}~\bibnamefont {Lee}}, \bibinfo {author} {\bibfnamefont
  {Y.}~\bibnamefont {Ronen}}, \bibinfo {author} {\bibfnamefont
  {H.}~\bibnamefont {Yoo}}, \bibinfo {author} {\bibfnamefont {D.}~\bibnamefont
  {Najafabadi}}, \bibinfo {author} {\bibfnamefont {K.}~\bibnamefont
  {Watanabe}}, \bibinfo {author} {\bibfnamefont {T.}~\bibnamefont {Taniguchi}},
  \bibinfo {author} {\bibfnamefont {A.}~\bibnamefont {Vishwanath}}, \ and\
  \bibinfo {author} {\bibfnamefont {P.}~\bibnamefont {Kim}},\ }\href {\doibase
  https://doi.org/10.1038/s41586-020-2458-7} {\bibfield  {journal} {\bibinfo
  {journal} {Nature}\ }\textbf {\bibinfo {volume} {583}},\ \bibinfo {pages}
  {221} (\bibinfo {year} {2020})}\BibitemShut {NoStop}%
\bibitem [{\citenamefont {Haddadi}\ \emph {et~al.}(2020)\citenamefont
  {Haddadi}, \citenamefont {Wu}, \citenamefont {Kruchkov},\ and\ \citenamefont
  {Yazyev}}]{doi:10.1021/acs.nanolett.9b05117}%
  \BibitemOpen
  \bibfield  {author} {\bibinfo {author} {\bibfnamefont {F.}~\bibnamefont
  {Haddadi}}, \bibinfo {author} {\bibfnamefont {Q.}~\bibnamefont {Wu}},
  \bibinfo {author} {\bibfnamefont {A.~J.}\ \bibnamefont {Kruchkov}}, \ and\
  \bibinfo {author} {\bibfnamefont {O.~V.}\ \bibnamefont {Yazyev}},\ }\href
  {\doibase 10.1021/acs.nanolett.9b05117} {\bibfield  {journal} {\bibinfo
  {journal} {Nano Lett.}\ }\textbf {\bibinfo {volume} {20}},\ \bibinfo {pages}
  {2410} (\bibinfo {year} {2020})}\BibitemShut {NoStop}%
\bibitem [{\citenamefont {Shen}\ \emph {et~al.}(2020)\citenamefont {Shen},
  \citenamefont {Chu}, \citenamefont {Wu}, \citenamefont {Li}, \citenamefont
  {Wang}, \citenamefont {Zhao}, \citenamefont {Tang}, \citenamefont {Liu},
  \citenamefont {Tian}, \citenamefont {Watanabe}, \citenamefont {Taniguchi},
  \citenamefont {Yang}, \citenamefont {Meng}, \citenamefont {Shi},
  \citenamefont {Yazyev},\ and\ \citenamefont {Zhang}}]{zhangguangyu2019}%
  \BibitemOpen
  \bibfield  {author} {\bibinfo {author} {\bibfnamefont {C.}~\bibnamefont
  {Shen}}, \bibinfo {author} {\bibfnamefont {Y.}~\bibnamefont {Chu}}, \bibinfo
  {author} {\bibfnamefont {Q.}~\bibnamefont {Wu}}, \bibinfo {author}
  {\bibfnamefont {N.}~\bibnamefont {Li}}, \bibinfo {author} {\bibfnamefont
  {S.}~\bibnamefont {Wang}}, \bibinfo {author} {\bibfnamefont {Y.}~\bibnamefont
  {Zhao}}, \bibinfo {author} {\bibfnamefont {J.}~\bibnamefont {Tang}}, \bibinfo
  {author} {\bibfnamefont {J.}~\bibnamefont {Liu}}, \bibinfo {author}
  {\bibfnamefont {J.}~\bibnamefont {Tian}}, \bibinfo {author} {\bibfnamefont
  {K.}~\bibnamefont {Watanabe}}, \bibinfo {author} {\bibfnamefont
  {T.}~\bibnamefont {Taniguchi}}, \bibinfo {author} {\bibfnamefont
  {R.}~\bibnamefont {Yang}}, \bibinfo {author} {\bibfnamefont {Z.~Y.}\
  \bibnamefont {Meng}}, \bibinfo {author} {\bibfnamefont {D.}~\bibnamefont
  {Shi}}, \bibinfo {author} {\bibfnamefont {O.~V.}\ \bibnamefont {Yazyev}}, \
  and\ \bibinfo {author} {\bibfnamefont {G.}~\bibnamefont {Zhang}},\ }\href
  {\doibase https://doi.org/10.1038/s41567-020-0825-9} {\bibfield  {journal}
  {\bibinfo  {journal} {Nat. Phys}\ }\textbf {\bibinfo {volume} {16}},\
  \bibinfo {pages} {520} (\bibinfo {year} {2020})}\BibitemShut {NoStop}%
\bibitem [{\citenamefont {Samajdar}\ \emph {et~al.}(2021)\citenamefont
  {Samajdar}, \citenamefont {Scheurer}, \citenamefont {Turkel}, \citenamefont
  {Rubio-Verd{\'{u}}}, \citenamefont {Pasupathy}, \citenamefont {Venderbos},\
  and\ \citenamefont {Fernandes}}]{Samajdar_2021}%
  \BibitemOpen
  \bibfield  {author} {\bibinfo {author} {\bibfnamefont {R.}~\bibnamefont
  {Samajdar}}, \bibinfo {author} {\bibfnamefont {M.~S.}\ \bibnamefont
  {Scheurer}}, \bibinfo {author} {\bibfnamefont {S.}~\bibnamefont {Turkel}},
  \bibinfo {author} {\bibfnamefont {C.}~\bibnamefont {Rubio-Verd{\'{u}}}},
  \bibinfo {author} {\bibfnamefont {A.~N.}\ \bibnamefont {Pasupathy}}, \bibinfo
  {author} {\bibfnamefont {J.~W.~F.}\ \bibnamefont {Venderbos}}, \ and\
  \bibinfo {author} {\bibfnamefont {R.~M.}\ \bibnamefont {Fernandes}},\ }\href
  {https://iopscience.iop.org/article/10.1088/2053-1583/abfcd6} {\bibfield
  {journal} {\bibinfo  {journal} {2D Mater.}\ }\textbf {\bibinfo {volume}
  {8}},\ \bibinfo {pages} {034005} (\bibinfo {year} {2021})}\BibitemShut
  {NoStop}%
\bibitem [{\citenamefont {Zhou}\ \emph {et~al.}(2022)\citenamefont {Zhou},
  \citenamefont {Holleis}, \citenamefont {Saito}, \citenamefont {Cohen},
  \citenamefont {Huynh}, \citenamefont {Patterson}, \citenamefont {Yang},
  \citenamefont {Taniguchi}, \citenamefont {Watanabe},\ and\ \citenamefont
  {Young}}]{zhou2022science}%
  \BibitemOpen
  \bibfield  {author} {\bibinfo {author} {\bibfnamefont {H.}~\bibnamefont
  {Zhou}}, \bibinfo {author} {\bibfnamefont {L.}~\bibnamefont {Holleis}},
  \bibinfo {author} {\bibfnamefont {Y.}~\bibnamefont {Saito}}, \bibinfo
  {author} {\bibfnamefont {L.}~\bibnamefont {Cohen}}, \bibinfo {author}
  {\bibfnamefont {W.}~\bibnamefont {Huynh}}, \bibinfo {author} {\bibfnamefont
  {C.}~\bibnamefont {Patterson}}, \bibinfo {author} {\bibfnamefont
  {F.}~\bibnamefont {Yang}}, \bibinfo {author} {\bibfnamefont {T.}~\bibnamefont
  {Taniguchi}}, \bibinfo {author} {\bibfnamefont {K.}~\bibnamefont {Watanabe}},
  \ and\ \bibinfo {author} {\bibfnamefont {A.}~\bibnamefont {Young}},\ }\href
  {\doibase 10.1126/science.abm8386} {\bibfield  {journal} {\bibinfo  {journal}
  {Science}\ }\textbf {\bibinfo {volume} {375}},\ \bibinfo {pages} {774}
  (\bibinfo {year} {2022})}\BibitemShut {NoStop}%
\bibitem [{\citenamefont {{de la Barrera}}\ \emph {et~al.}(2022)\citenamefont
  {{de la Barrera}}, \citenamefont {{Aronson}}, \citenamefont {{Zheng}},
  \citenamefont {{Watanabe}}, \citenamefont {{Taniguchi}}, \citenamefont
  {{Ma}}, \citenamefont {{Jarillo-Herrero}},\ and\ \citenamefont
  {{Ashoori}}}]{blg2022natphy}%
  \BibitemOpen
  \bibfield  {author} {\bibinfo {author} {\bibfnamefont {S.~C.}\ \bibnamefont
  {{de la Barrera}}}, \bibinfo {author} {\bibfnamefont {S.}~\bibnamefont
  {{Aronson}}}, \bibinfo {author} {\bibfnamefont {Z.}~\bibnamefont {{Zheng}}},
  \bibinfo {author} {\bibfnamefont {K.}~\bibnamefont {{Watanabe}}}, \bibinfo
  {author} {\bibfnamefont {T.}~\bibnamefont {{Taniguchi}}}, \bibinfo {author}
  {\bibfnamefont {Q.}~\bibnamefont {{Ma}}}, \bibinfo {author} {\bibfnamefont
  {P.}~\bibnamefont {{Jarillo-Herrero}}}, \ and\ \bibinfo {author}
  {\bibfnamefont {R.}~\bibnamefont {{Ashoori}}},\ }\href {\doibase
  10.1038/s41567-022-01616-w} {\bibfield  {journal} {\bibinfo  {journal} {Nat.
  Phys.}\ }\textbf {\bibinfo {volume} {18}},\ \bibinfo {pages} {771} (\bibinfo
  {year} {2022})}\BibitemShut {NoStop}%
\bibitem [{\citenamefont {Seiler}\ \emph {et~al.}(2022)\citenamefont {Seiler},
  \citenamefont {Geisenhof}, \citenamefont {Winterer}, \citenamefont
  {Watanabe}, \citenamefont {Taniguchi}, \citenamefont {Xu}, \citenamefont
  {Zhang},\ and\ \citenamefont {Weitz}}]{blg2022nature}%
  \BibitemOpen
  \bibfield  {author} {\bibinfo {author} {\bibfnamefont {A.}~\bibnamefont
  {Seiler}}, \bibinfo {author} {\bibfnamefont {F.}~\bibnamefont {Geisenhof}},
  \bibinfo {author} {\bibfnamefont {F.}~\bibnamefont {Winterer}}, \bibinfo
  {author} {\bibfnamefont {K.}~\bibnamefont {Watanabe}}, \bibinfo {author}
  {\bibfnamefont {T.}~\bibnamefont {Taniguchi}}, \bibinfo {author}
  {\bibfnamefont {T.}~\bibnamefont {Xu}}, \bibinfo {author} {\bibfnamefont
  {F.}~\bibnamefont {Zhang}}, \ and\ \bibinfo {author} {\bibfnamefont
  {R.}~\bibnamefont {Weitz}},\ }\href {\doibase 10.1038/s41586-022-04937-1}
  {\bibfield  {journal} {\bibinfo  {journal} {Nature}\ }\textbf {\bibinfo
  {volume} {608}},\ \bibinfo {pages} {298} (\bibinfo {year}
  {2022})}\BibitemShut {NoStop}%
\end{thebibliography}%
	
\end{document}